\documentclass[twocolumn]{aastex63}
\usepackage{amsmath}
\usepackage{graphicx}
\usepackage{natbib}
\usepackage[scaled]{helvet}
\usepackage{epsfig}
\usepackage{url}
\bibpunct{(}{)}{;}{a}{}{,}
\newcommand{\kms}{\,km\,s$^{-1}$}
\usepackage{textcomp}
\usepackage{mathpazo}
\usepackage{xspace}
\usepackage{hyperref}
\usepackage{apjfonts}
\usepackage{mathrsfs}
\usepackage{verbatim}

\usepackage{color}
\usepackage[normalem]{ulem}

\newcommand{\bjdtdb}{\ensuremath{\rm {BJD_{TDB}}}}
\newcommand{\feh}{\ensuremath{\left[{\rm Fe}/{\rm H}\right]}}

\newcommand{\teff}{\ensuremath{T_{\rm eff}}\xspace}
\newcommand{\logg}{\ensuremath{\log g}}

\newcommand{\msun}{\ensuremath{\,M_\Sun}}
\newcommand{\rsun}{\ensuremath{\,R_\Sun}}
\newcommand{\lsun}{\ensuremath{\,L_\Sun}}
\newcommand{\mj}{\ensuremath{\,M_{\rm J}}}
\newcommand{\rj}{\ensuremath{\,R_{\rm J}}}

\newcommand{\vsini}{\ensuremath{v\sin{I_*}}}
\newcommand{\ms}{\,m\,s$^{-1}$}
\newcommand{\mstar}{\ensuremath{M_{*}}}

\newcommand{\be}{\begin{equation}}
\newcommand{\ee}{\end{equation}}

\newcommand{\tess}{{\it TESS}}
\newcommand{\TESS}{{\it TESS}}

\usepackage{lineno}

\begin{document}

\title{TESS Delivers Five New Hot Giant Planets Orbiting Bright Stars from the Full Frame Images}

\newcommand{\cfa}{Center for Astrophysics \textbar \ Harvard \& Smithsonian, 60 Garden St, Cambridge, MA 02138, USA}
\newcommand{\msu}{Department of Physics and Astronomy, Michigan State University, East Lansing, MI 48824, USA}
\newcommand{\umich}{Astronomy Department, University of Michigan, 1085 S University Avenue, Ann Arbor, MI 48109, USA}
\newcommand{\utaustin}{Department of Astronomy, The University of Texas at Austin, Austin, TX 78712, USA}
\newcommand{\MIT}{Department of Physics and Kavli Institute for Astrophysics and Space Research, Massachusetts Institute of Technology, Cambridge, MA 02139, USA}
\newcommand{\MITEPS}{Department of Earth, Atmospheric and Planetary Sciences, Massachusetts Institute of Technology,  Cambridge,  MA 02139, USA}
\newcommand{\uflorida}{Department of Astronomy, University of Florida, 211 Bryant Space Science Center, Gainesville, FL, 32611, USA}
\newcommand{\riverside}{Department of Earth and Planetary Sciences, University of California, Riverside, CA 92521, USA}
\newcommand{\usq}{Centre for Astrophysics, University of Southern Queensland, West Street, Toowoomba, QLD 4350, Australia}
\newcommand{\ames}{NASA Ames Research Center, Moffett Field, CA, 94035, USA}
\newcommand{\geneva}{Geneva Observatory, University of Geneva, Chemin des Mailettes 51, 1290 Versoix, Switzerland}
\newcommand{\uw}{Astronomy Department, University of Washington, Seattle, WA 98195 USA}
\newcommand{\warwick}{Department of Physics, University of Warwick, Gibbet Hill Road, Coventry CV4 7AL, UK}
\newcommand{\warwickceh}{Centre for Exoplanets and Habitability, University of Warwick, Gibbet Hill Road, Coventry CV4 7AL, UK}
\newcommand{\princeton}{Department of Astrophysical Sciences, Princeton University, 4 Ivy Lane, Princeton, NJ, 08544, USA}
\newcommand{\liege}{Space Sciences, Technologies and Astrophysics Research (STAR) Institute, Universit\'e de Li\`ege, 19C All\'ee du 6 Ao\^ut, 4000 Li\`ege, Belgium}
\newcommand{\vanderbilt}{Department of Physics and Astronomy, Vanderbilt University, Nashville, TN 37235, USA}
\newcommand{\fisk}{Department of Physics, Fisk University, 1000 17th Avenue North, Nashville, TN 37208, USA}
\newcommand{\columbia}{Department of Astronomy, Columbia University, 550 West 120th Street, New York, NY 10027, USA}
\newcommand{\toronto}{Dunlap Institute for Astronomy and Astrophysics, University of Toronto, Ontario M5S 3H4, Canada}
\newcommand{\unc}{Department of Physics and Astronomy, University of North Carolina at Chapel Hill, Chapel Hill, NC 27599, USA}
\newcommand{\iac}{Instituto de Astrof\'isica de Canarias (IAC), E-38205 La Laguna, Tenerife, Spain}
\newcommand{\lalaguna}{Departamento de Astrof\'isica, Universidad de La Laguna (ULL), E-38206 La Laguna, Tenerife, Spain}
\newcommand{\louisville}{Department of Physics and Astronomy, University of Louisville, Louisville, KY 40292, USA}
\newcommand{\aavso}{American Association of Variable Star Observers, 49 Bay State Road, Cambridge, MA 02138, USA}
\newcommand{\utokyo}{The University of Tokyo, 7-3-1 Hongo, Bunky\={o}, Tokyo 113-8654, Japan}
\newcommand{\naoj}{National Astronomical Observatory of Japan, 2-21-1 Osawa, Mitaka, Tokyo 181-8588, Japan}
\newcommand{\jstpresto}{JST, PRESTO, 7-3-1 Hongo, Bunkyo-ku, Tokyo 113-0033, Japan}
\newcommand{\astrobiojapan}{Astrobiology Center, 2-21-1 Osawa, Mitaka, Tokyo 181-8588, Japan}
\newcommand{\ctio}{Cerro Tololo Inter-American Observatory, Casilla 603, La Serena, Chile}
\newcommand{\noirlab}{NOIRLab/Cerro Tololo Inter-American Observatory, Casilla 603, La Serena, Chile}
\newcommand{\nexsci}{Caltech IPAC -- NASA Exoplanet Science Institute 1200 E. California Ave, Pasadena, CA 91125, USA}
\newcommand{\ucsc}{Department of Astronomy and Astrophysics, University of
California, Santa Cruz, CA 95064, USA}
\newcommand{\gsfc}{Exoplanets and Stellar Astrophysics Laboratory, Code 667, NASA Goddard Space Flight Center, Greenbelt, MD 20771, USA}
\newcommand{\sgtinc}{SGT, Inc./NASA AMES Research Center, Mailstop 269-3, Bldg T35C, P.O. Box 1, Moffett Field, CA 94035, USA}
\newcommand{\chile}{Center of Astro-Engineering UC, Pontificia Universidad Cat\'olica de Chile, Av. Vicu\~{n}a Mackenna 4860, 7820436 Macul, Santiago, Chile}
\newcommand{\Pontificia}{Facultad de Ingeniería y Ciencias, Universidad Adolfo Ib\'a\~nez, Av. Diagonal las Torres 2640, Pe\~nalol\'en, Santiago, Chile}
\newcommand{\Millennium}{Millennium Institute for Astrophysics, Chile}
\newcommand{\maxplank}{Max-Planck-Institut f\"ur Astronomie, K\"onigstuhl 17, Heidelberg 69117, Germany}
\newcommand{\utdallas}{Department of Physics, The University of Texas at Dallas, 800 West
Campbell Road, Richardson, TX 75080-3021 USA}
\newcommand{\MauryLewin}{Maury Lewin Astronomical Observatory, Glendora, CA 91741, USA}
\newcommand{\umbc}{University of Maryland, Baltimore County, 1000 Hilltop Circle, Baltimore, MD 21250, USA}
\newcommand{\osu}{Department of Astronomy, The Ohio State University, 140 West 18th Avenue, Columbus, OH 43210, USA}
\newcommand{\MITAA}{Department of Aeronautics and Astronautics, MIT, 77 Massachusetts Avenue, Cambridge, MA 02139, USA}
\newcommand{\openu}{School of Physical Sciences, The Open University, Milton Keynes MK7 6AA, UK}
\newcommand{\swarthmore}{Department of Physics and Astronomy, Swarthmore College, Swarthmore, PA 19081, USA}
\newcommand{\seti}{SETI Institute, Mountain View, CA 94043, USA}
\newcommand{\lehigh}{Department of Physics, Lehigh University, 16 Memorial Drive East, Bethlehem, PA 18015, USA}
\newcommand{\utah}{Department of Physics and Astronomy, University of Utah, 115 South 1400 East, Salt Lake City, UT 84112, USA}
\newcommand{\USNA}{Department of Physics, United States Naval Academy, 572C Holloway Rd., Annapolis, MD 21402, USA}
\newcommand{\DTM}{Department of Terrestrial Magnetism, Carnegie Institution for Science, 5241 Broad Branch Road, NW, Washington, DC 20015, USA}
\newcommand{\UPenn}{The University of Pennsylvania, Department of Physics and Astronomy, Philadelphia, PA, 19104, USA}
\newcommand{\montana}{Department of Physics and Astronomy, University of Montana, 32 Campus Drive, No. 1080, Missoula, MT 59812 USA}
\newcommand{\psu}{Department of Astronomy \& Astrophysics, The Pennsylvania State University, 525 Davey Lab, University Park, PA 16802, USA}
\newcommand{\psust}{Center for Exoplanets and Habitable Worlds, The Pennsylvania State University, 525 Davey Lab, University Park, PA 16802, USA}
\newcommand{\Kutztown}{Department of Physical Sciences, Kutztown University, Kutztown, PA 19530, USA}
\newcommand{\udel}{Department of Physics \& Astronomy, University of Delaware, Newark, DE 19716, USA}
\newcommand{\Westminster}{Department of Physics, Westminster College, New Wilmington, PA 16172}
\newcommand{\steward}{Department of Astronomy and Steward Observatory, University of Arizona, Tucson, AZ 85721, USA}
\newcommand{\saao}{South African Astronomical Observatory, PO Box 9, Observatory, 7935, Cape Town, South Africa}
\newcommand{\salt}{Southern African Large Telescope, PO Box 9, Observatory, 7935, Cape Town, South Africa}
\newcommand{\ssl}{Societ\`{a} Astronomica Lunae, Italy}
\newcommand{\spot}{Spot Observatory, Nashville, TN 37206, USA}
\newcommand{\txamGP}{George P.\ and Cynthia Woods Mitchell Institute for Fundamental Physics and Astronomy, Texas A\&M University, College Station, TX77843 USA}
\newcommand{\txam}{Department of Physics and Astronomy, Texas A\&M university, College Station, TX 77843 USA}
\newcommand{\wellesley}{Department of Astronomy, Wellesley College, Wellesley, MA 02481, USA}
\newcommand{\Wesleyan}{Astronomy Department and Van Vleck Observatory, Wesleyan University, Middletown, CT 06459, USA}
\newcommand{\inaf}{INAF -- Osservatorio Astronomico di Padova, Vicolo dell'Osservatorio 5, I-35122, Padova, Italy}
\newcommand{\byu}{Department of Physics and Astronomy, Brigham Young University, Provo, UT 84602, USA}
\newcommand{\Hazelwood}{Hazelwood Observatory, Churchill, Victoria, Australia}
\newcommand{\pest}{Perth Exoplanet Survey Telescope}
\newcommand{\Winer}{Winer Observatory, PO Box 797, Sonoita, AZ 85637, USA}
\newcommand{\icpo}{Ivan Curtis Private Observatory}
\newcommand{\elsauce}{El Sauce Observatory, Chile}
\newcommand{\crow}{Atalaia Group \& CROW Observatory, Portalegre, Portugal}
\newcommand{\dfus}{Dipartimento di Fisica ``E.R.Caianiello'', Universit\`a di Salerno, Via Giovanni Paolo II 132, Fisciano 84084, Italy}
\newcommand{\indfn}{Istituto Nazionale di Fisica Nucleare, Napoli, Italy}
\newcommand{\sotes}{Gabriel Murawski Private Observatory (SOTES)}
\newcommand{\lco}{Las Cumbres Observatory Global Telescope, 6740 Cortona Dr., Suite 102, Goleta, CA 93111, USA}
\newcommand{\ucsb}{Department of Physics, University of California, Santa Barbara, CA 93106-9530, USA}
\newcommand{\yale}{Department of Astronomy, Yale University, 52 Hillhouse Avenue, New Haven, CT 06511, USA}
\newcommand{\eso}{European Southern Observatory, Alonso de C\'ordova 3107, Vitacura, Casilla 19001, Santiago, Chile}
\newcommand{\stsci}{Space Telescope Science Institute, Baltimore, MD 21218, USA}
\newcommand{\keele}{Astrophysics Group, Keele University, Staffordshire ST5 5BG, UK}
\newcommand{\gsfcsellers}{GSFC Sellers Exoplanet Environments Collaboration, NASA Goddard Space Flight Center, Greenbelt, MD 20771 }
\newcommand{\usno}{U.S. Naval Observatory, Washington, DC 20392, USA}
\newcommand{\kansas}{Department of Physics and Astronomy, University of Kansas, 1251 Wescoe Hall Dr., Lawrence, KS 66045, USA}
\newcommand{\gmu}{George Mason University, 4400 University Drive MS 3F3, Fairfax, VA 22030, USA}
\newcommand{\unsw}{Exoplanetary Science at UNSW, School of Physics, UNSW Sydney, NSW 2052, Australia}
\newcommand{\sifa}{School of Physics, Sydney Institute for Astronomy (SIfA), The University of Sydney, NSW 2006, Australia}
\newcommand{\nanjing}{School of Astronomy and Space Science, Key Laboratory of Modern Astronomy and Astrophysics in Ministry of Education, Nanjing University, Nanjing 210046, Jiangsu, China}
\newcommand{\berkely}{Department of Astronomy, University of California Berkeley, Berkeley, CA 94720-3411, USA}
\newcommand{\bhicfa}{Black Hole Initiative at Harvard University, 20 Garden Street, Cambridge, MA 02138, USA}
\newcommand{\Silesian}{Department of Electronics, Electronical Engineering and Microelectronics, Silesian University of Techhnology Akademicka 16, 44-100 Gliwice, Poland}
\newcommand{\Patashnick}{Patashnick Voorheesville Observatory, Voorheesville, NY 12186, USA}
\newcommand{\austincollege}{Physics Department, Austin College, 900 North Grand Avenue, Sherman TX 75090, USA}
\newcommand{\Tsinghua}{Department of Astronomy, Tsinghua University, Beijing 100084, China}
\newcommand{\Tsinghuaschool}{Tsinghua International School, Beijing 100084, China}
\newcommand{\chinaNAO}{National Astronomical Observatories, Chinese Academy of Sciences, 20A Datun Road, Chaoyang District, Beijing 100012, China}
\newcommand{\Tautenburg}{Th{\"u}ringer Landessternwarte Tautenburg, Sternwarte 5, 07778 Tautenburg, Germany}
\newcommand{\brierfield}{Brierfield Observatory, New South Wales, Australia}
\newcommand{\Indiana}{Indiana University Department of Astronomy, SW319, 727 E 3rd Street, Bloomington, IN 47405 USA}
\newcommand{\wisconsin}{Department of Astronomy, University of Wisconsin-Madison, Madison, WI 53706, USA}
\newcommand{\protologic}{Proto-Logic Consulting LLC, Washington, DC 20009, USA}
\newcommand{\ASTRAVEO}{ASTRAVEO LLC, PO Box 1668, MA 01931}
\newcommand{\TJHS}{Thomas Jefferson High School, 6560 Braddock Rd, Alexandria, VA 22312 USA}
\newcommand{\ucatchile}{Instituto de Astrof\'isica, Facultad de F\'isica, Pontificia Universidad Cat\'olica de Chile}
\newcommand{\lasa}{Liberal Arts and Science Academy, Austin, Texas 78724, USA}
\newcommand{\gemini}{Gemini Observatory/NSF’s NOIRLab, 670 N. A’ohoku Place, Hilo, HI, 96720, USA}
\newcommand{\umd}{Department of Astronomy, University of Maryland, College Park, College Park, MD}

\newcommand{\eberly}{\altaffiliation{Eberly Research Fellow}}
\newcommand{\torres}{\altaffiliation{Juan Carlos Torres Fellow}}
\newcommand{\sagan}{\altaffiliation{NASA Sagan Fellow}}
\newcommand{\bernoulli}{\altaffiliation{Bernoulli fellow}}
\newcommand{\gruber}{\altaffiliation{Gruber fellow}}
\newcommand{\kavli}{\altaffiliation{Kavli Fellow}}
\newcommand{\peg}{\altaffiliation{51 Pegasi b Fellow}}
\newcommand{\pappalardo}{\altaffiliation{Pappalardo Fellow}}
\newcommand{\hubble}{\altaffiliation{NASA Hubble Fellow}}
\newcommand{\nsf}{\altaffiliation{National Science Foundation Graduate Research Fellow}}

\correspondingauthor{Joseph E. Rodriguez} 
\email{jrod@msu.edu}

\author[0000-0001-8812-0565]{Joseph E. Rodriguez} 
\affiliation{\msu}
\affiliation{\cfa}

\author[0000-0002-8964-8377]{Samuel N. Quinn} 
\affiliation{\cfa}

\author[0000-0002-4891-3517]{George Zhou} 
\affiliation{\cfa}

\author[0000-0001-7246-5438]{Andrew Vanderburg} 
\affiliation{\wisconsin}

\author[0000-0002-5254-2499]{Louise D. Nielsen}
\affiliation{\geneva}

\author[0000-0001-9957-9304]{Robert A. Wittenmyer} 
\affil{\usq}

\author[0000-0002-9158-7315]{Rafael Brahm} 
\affiliation{\Pontificia}
\affiliation{\Millennium}

\author[0000-0002-5005-1215]{Phillip A.\ Reed} 
\affiliation{\Kutztown}

\author[0000-0003-0918-7484]{Chelsea X. Huang} 
\torres
\affiliation{\MIT}

\author{Sydney Vach} 
\affiliation{\cfa}

% ...

\shorttitle{Everybody Gets a Giant Planet!}
\shortauthors{Rodriguez et al.}

\begin{abstract}
We present the discovery and characterization of five hot and warm Jupiters --- TOI-628 b (TIC 281408474; HD 288842), TOI-640 b (TIC 147977348), TOI-1333 b (TIC 395171208, BD+47 3521A), TOI-1478 b (TIC 409794137), and TOI-1601 b (TIC 139375960) --- based on data from NASA’s {\it Transiting Exoplanet Survey Satellite (TESS)}. The five planets were identified from the full frame images and were confirmed through a series of photometric and spectroscopic follow-up observations by the {\it TESS} Follow-up Observing Program (TFOP) Working Group. The planets are all Jovian size (R$_{\rm P}$ = 1.01-1.77 R$_{\rm J}$) and have masses that range from 0.85 to 6.33 M$_{\rm J}$. The host stars of these systems have F and G spectral types (5595 $\le$ T$_{\rm eff}$ $\le$ 6460 K) and are all relatively bright (9.5 $<V<$ 10.8, 8.2 $<K<$ 9.3) making them well-suited for future detailed characterization efforts. Three of the systems in our sample (TOI-640 b, TOI-1333 b, and  TOI-1601 b) orbit subgiant host stars ($\log$ g $<$4.1). TOI-640 b is one of only three known hot Jupiters to have a highly inflated radius (R$_{\rm P}$ > 1.7R$_{\rm J}$, possibly a result of its host star's evolution) and resides on an orbit with a period longer than 5 days. TOI-628 b is the most massive hot Jupiter discovered to date by {\it TESS} with a measured mass of $6.31^{+0.28}_{-0.30}$ M$_{\rm J}$ and a statistically significant, non-zero orbital eccentricity of e = $0.074^{+0.021}_{-0.022}$. This planet would not have had enough time to circularize through tidal forces from our analysis, suggesting that it might be remnant eccentricity from its migration. The longest period planet in this sample, TOI-1478 b (P = 10.18 days), is a warm Jupiter in a circular orbit around a near-Solar analogue. NASA's {\it TESS} mission is continuing to increase the sample of well-characterized hot and warm Jupiters, complementing its primary mission goals. 
\end{abstract}

\section{Introduction}


The discovery of hot Jupiters, combined with the assumption that gas giant planets must form at separations from their host star similar to our own giant planets, indicated that giant planets likely undergo large-scale migration from their formation locations.  Various mechanisms have been proposed to emplace giant planets into very short-period orbits (\citealp{Goldreich:1980, Lin:1986, Lin:1996}; see \citealp{Dawson:2018} for a detailed review). However, it is not clear which of these mechanisms (if any) are dominant, govern this migration, or whether hot Jupiters {\it can} form in situ \citep{Batygin:2016}, obviating the need for large-scale migration. One possibility is that giant planets migrate slowly and smoothly within the circumstellar gas-dust disk, resulting in well-aligned, nearly circular orbits \citep{Dangelo:2003}. It is also thought that planetary migration may be heavily influenced by gravitational interactions with other bodies within the system.  These interactions result in highly eccentric and misaligned orbits (relative to the rotation axis of the star), and are typically referred to as a type of ``High Eccentricity Migration" (HEM) or  ``Kozai-Lidov" \citep{Kozai:1962, Lidov:1962, Rasio:1996, Wu:2003, Fabrycky:2007, Nagasawa:2011, Wu:2011, Naoz:2016}. For short-period hot Jupiters, with periods less than about 5 days, the orbits will circularize in only a few billion years, erasing the evidence of HEM. Additionally, these interactions can cause misalignments in the planet's orbital plane (relative to the original disk plane), that can remain present for much longer, and such misalignments can be detected through Doppler spectroscopy, using observations of the Rossiter McLauglin effect \citep{Rossiter:1924, McLaughlin:1924} or Doppler tomography \citep[e.g.,][]{Miller:2010, Johnson:2014, Zhou:2016}.  Longer-period hot (P $\gtrsim$ 5 days) and warm (P$>$10 days) Jupiters experience smaller tidal forces, preserving their orbital eccentricities. It is likely that multiple mechanisms shape the short-period giant planet population, and studying these longer period hot Jupiters may give clues to their common evolutionary pathway.

\begin{figure*}[!ht]
\centering 
\includegraphics[trim = 0 0 0 0,width=\linewidth]{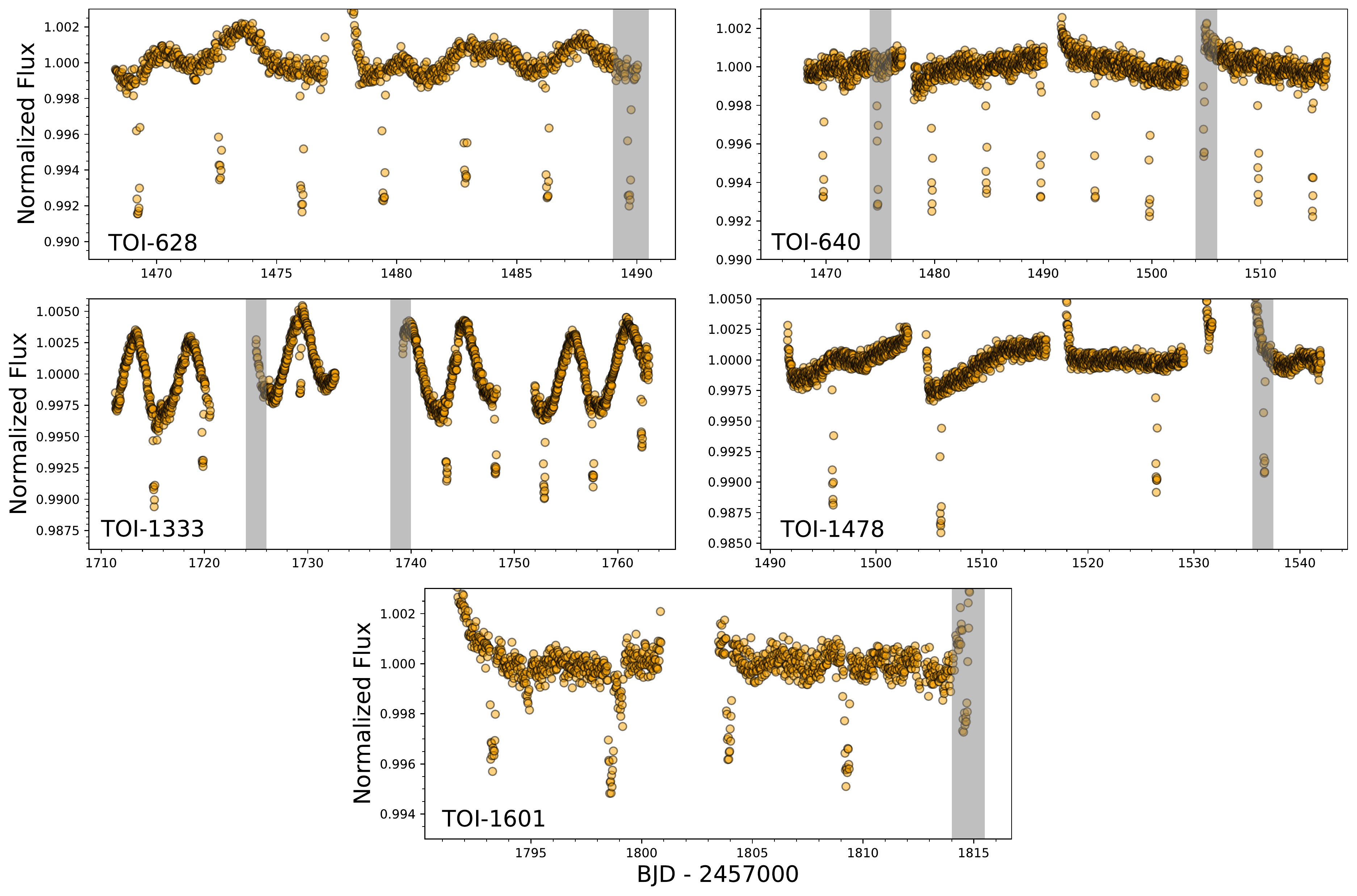}
\caption{The raw \tess\ QLP 30-minute light curves for (top-left) TOI-628, TOI-640 (top-right), TOI-1333 (middle-left), TOI-1478 (middle-right), and TOI-1601 (bottom). Transits highlighted in gray were excluded from the global fit since they were flagged as bad quality by the QLP pipeline \citep{Huang:QLP}. }
\label{fig:fullLCs}
\end{figure*}

While some ground-based transit surveys that were dedicated to discovering hot Jupiters had near 24-hour coverage \citep{Bakos:2013}, in general they struggled to discover planets with periods $\gtrsim$5 days due to the poor duty cycle from weather and only being able to observe at night \citep{Gaudi:2005}. Additionally, many of the first hot Jupiters discovered were assumed to reside in circular orbits when analyzing the observations, an assumption that may confuse current efforts to understand migration. Fortunately, NASA's Transiting Exoplanet Survey Satellite (\tess) mission was launched in April of 2018, and completed its primary 2-year long mission in July of 2020 \citep{Ricker:2015}. \tess\ was awarded a 27-month first extended mission in which it will not only re-observe some areas covered in the primary mission but also observe most of the ecliptic plane, nearly completing coverage of the entire sky. \tess\ has a minimum observing baseline of $\sim$27 days, and from recent occurrence rate studies, \tess\ planet searches will be mostly complete for hot Jupiters with periods $\lessapprox$10 days \citep{Zhou:2019}. Therefore, \tess\ provides a great resource for the discovery and confirmation of new longer-period hot and warm Jupiters (5$<P<$15 days) where eccentricities from migration would not be completely erased by tidal forces. \tess\ has already discovered a number of statistically significant hot Jupiters with highly eccentric (e $>$ 0.2) orbits like HD 2685 b \citep{Jones:2019}, TOI-172 b \citep{Rodriguez:2019},  TOI-150 b \citep{Kossakowski:2019}, TIC 237913194 b \citep{Schlecker:2020}, and TOI-559 b \citep{IkwutUkwa:2021}. Additionally, \tess\ recently confirmed that the hot Jupiter HD 118203 b, an RV-identified planet with high eccentricity discovered with the radial velocity method \citep{daSilva:2006}, transits its host star \citep{Pepper:2019}. 

\tess\ will also provide the ability to study hot Jupiter re-inflation since its high photometric precision will allow it to discover giant planets around larger, more evolved host stars. As a star evolves off the main sequence, the stellar irradiation received by warm Jupiters is similar to that of a hot Jupiter. Therefore, discovering gas giant planets orbiting evolved stars at longer periods (10s of days) can test whether this increased irradiation  causes the same inflation seen for short-period hot Jupiters \citep{Lopez:2016}. Most warm Jupiters orbiting main sequence stars show little to no inflation \citep{Demory:2011}, suggesting that this energy must be transferred deep into the planet's interior \citep{Liu:2008, Spiegel:2013} and as the star evolve, these warm Jupiters may re-inflate from the increased irradiation. Recent discoveries of hot Jupiters orbiting evolved stars are suggestive of reinflation \citep{Almenara:2015, Grunblatt:2016, Hartman:2016, Stevens:2017} and \tess\ has already found a few hot and warm Jupiters orbiting evolved stars \citep{Brahm:2019, Nielsen:2019, Huber:2019, Rodriguez:2019, Wang:2019, Sha:2020}.

In this paper we present the discovery of TOI-628 b, TOI-640 b, TOI-1333 b, TOI-1478 b, and TOI-1601 b, five new hot and warm Jupiters from NASA's {\it TESS} mission. All five planets were discovered from an analysis of the 30-minute cadence Full Frame Images (FFIs), and first identified as \tess\ Objects of Interest (TOIs) by the \TESS\ Science Office\footnote{\url{https://tess.mit.edu/toi-releases/toi-release-general/}}. These five new systems increase the known sample of well-characterized hot Jupiters, particularly those with longer orbital periods ($>$5 days). In \S\ref{sec:Obs} we present our time-series photometric and spectroscopic observations obtained by the {\it TESS} Follow-up Observing Program (TFOP) Working Group (WG) and describe the high spatial resolution imaging of all five targets, specifically on TOI-1333 and its two nearby companions. Our methodology for our global modeling using \texttt{EXOFASTv2} \citep{Eastman:2019} is summarized in \S\ref{sec:GlobalModel}. We place these systems in context with the known population of hot and warm Jupiters,  and discuss the impact of {\it TESS} in the discovery of giant planets in \S\ref{fig:discussion} and our conclusion are given in \S\ref{sec:conclusion}.

\begin{figure*}[ht]
	\centering\vspace{.0in}
	\includegraphics[width=0.9\linewidth, trim={0 0 0 0}, clip]{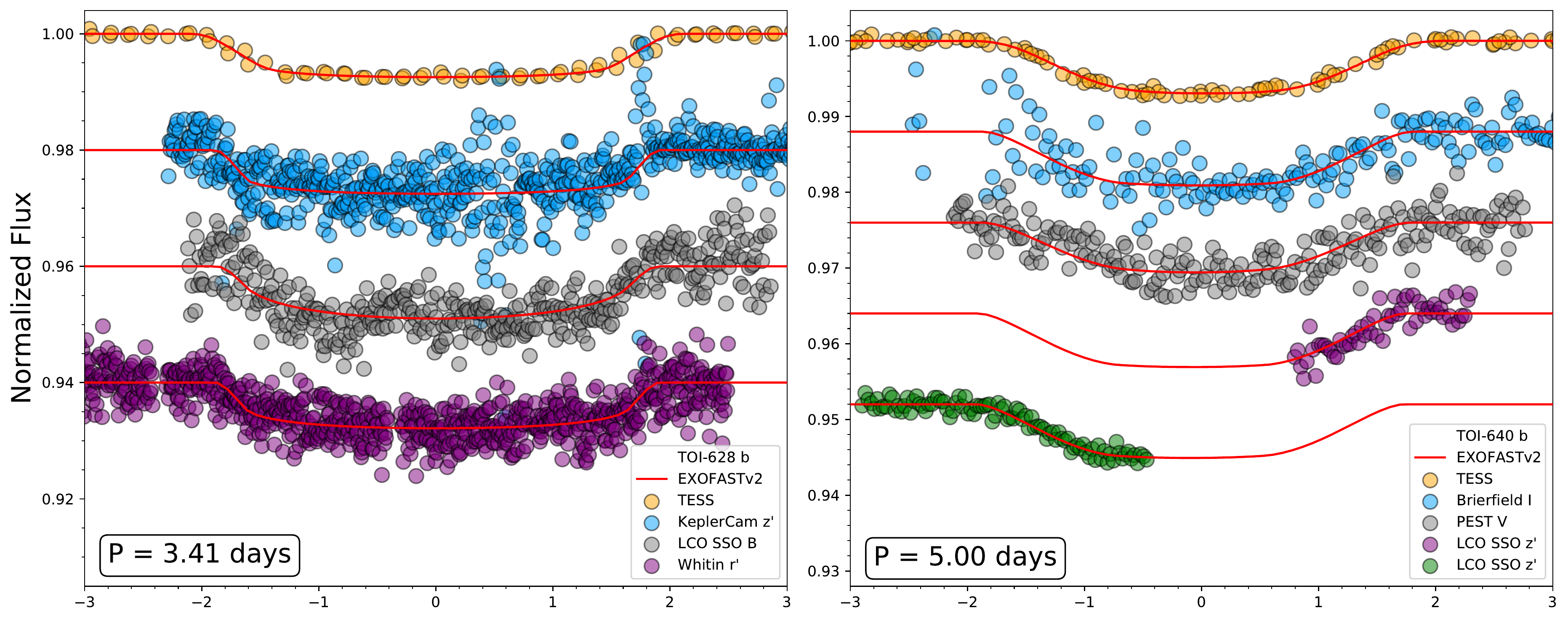}\\
	\includegraphics[width=0.9\linewidth, trim={0 0 0 0}, clip]{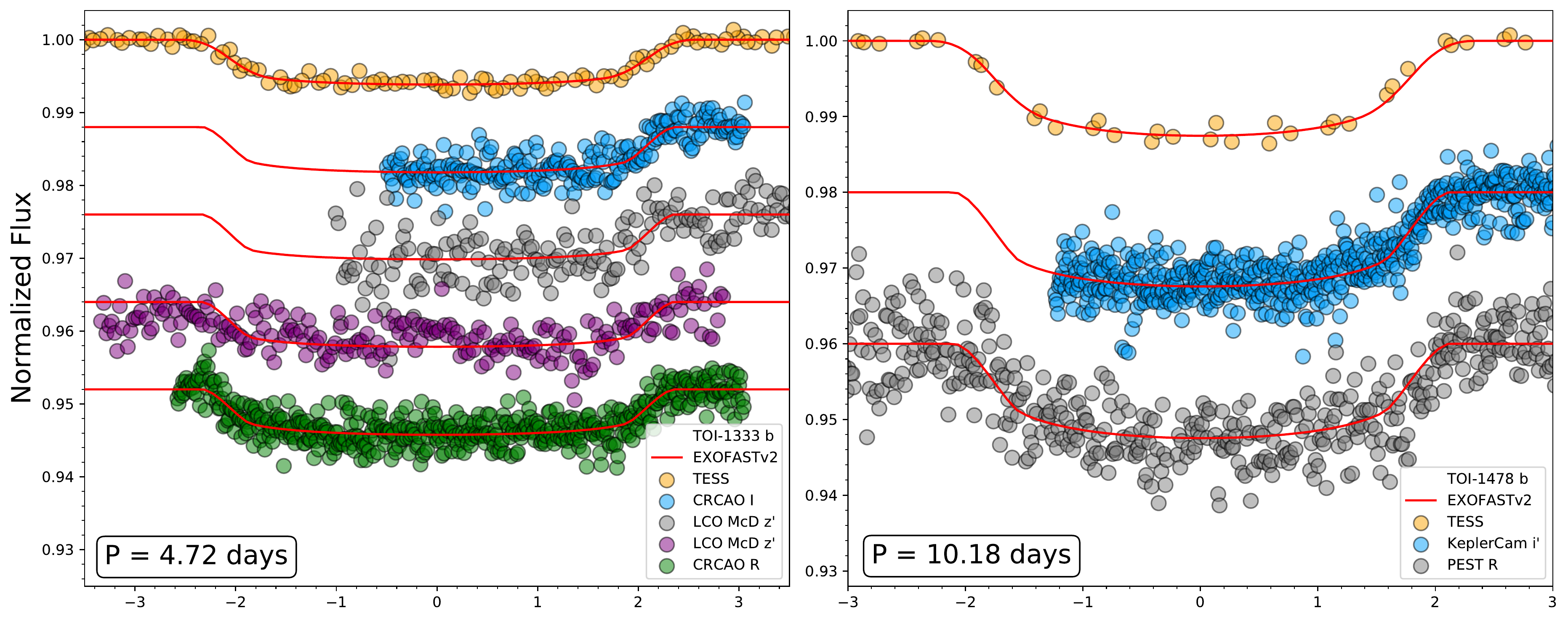}\\
	\includegraphics[width=0.5\linewidth, trim={0 0 0 0}, clip]{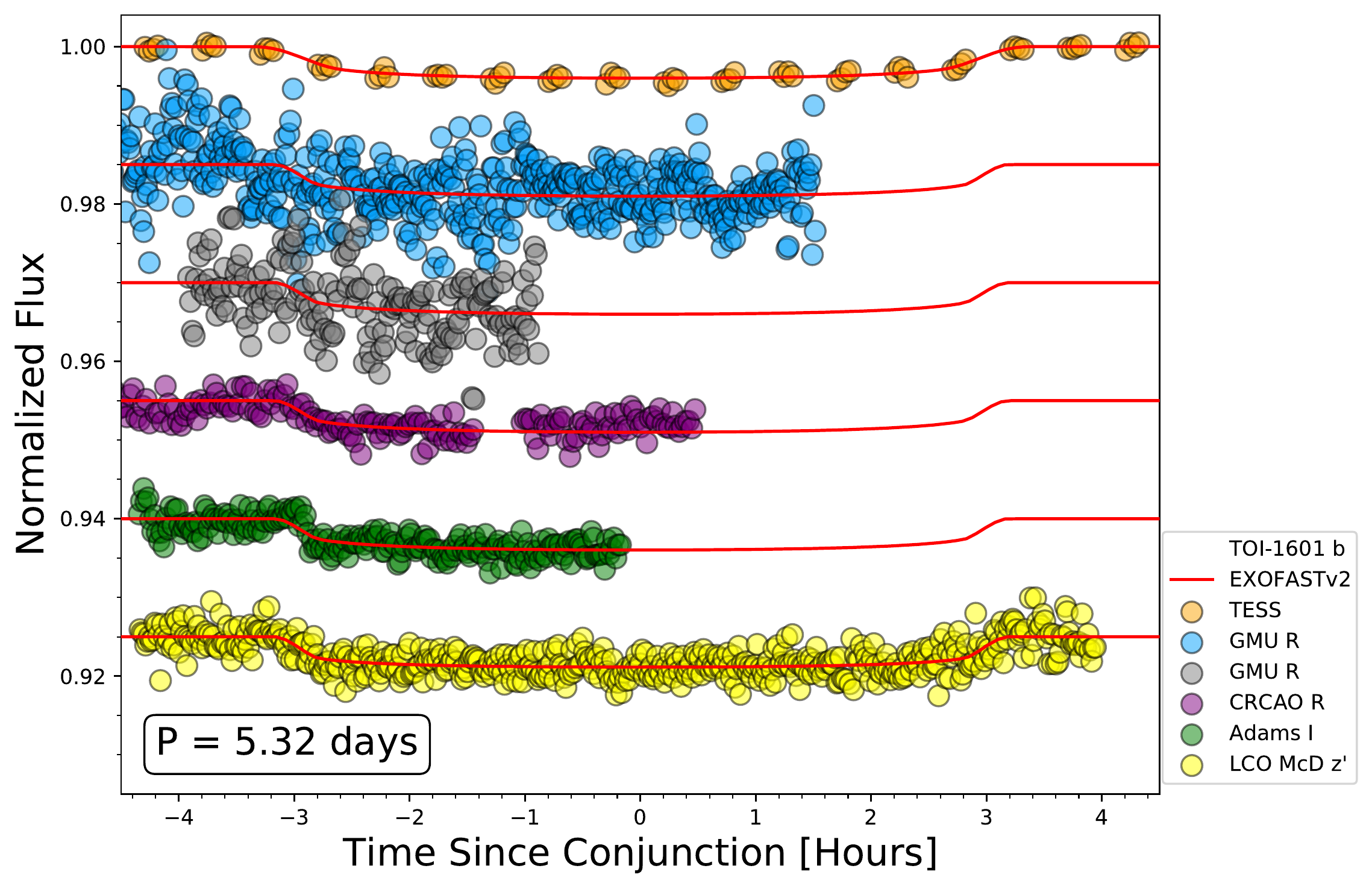}\\
	\caption{The \tess\ (orange) and TFOP SG1 follow-up transits of TOI-628 b (top-left), TOI-640 b (top-right), TOI-1333 b (middle-left), TOI-1478 b (middle-right), and TOI-1601 b (bottom). The \texttt{EXOFASTv2} model for each transit observation is shown by the red solid line.}
	\label{fig:transits} 
\end{figure*}

\section{Observations and Archival Data}
\label{sec:Obs}
We used a series photometric and spectroscopic observations to confirm and characterize the new systems, including high-spatial resolution imaging to rule out false positives, confirm them as bona-fide planets, and measure key parameters such as orbital eccentricity and the planet's density.

\subsection{{\it TESS} Photometry}
\label{sec:TESS}

The initial detection of the new planets  came from data collected by the \TESS\ mission. \TESS\ images the sky with a 24$^\circ$x96$^\circ$ field of view and observes the same stars for about a month before moving on to observe a different region. During its primary mission, \TESS\ saved and downloaded images of a smaller number of pre-selected stars every two minutes, while downloading co-added images of its entire field of view every 30 minutes. None of the planets described in this paper orbit stars that were pre-selected for two-minute cadence observations, so we use data from the 30-minute cadence FFIs. 

After the data were transferred from the orbiting spacecraft back to Earth, the FFIs were calibrated using the TICA software (Fausnaugh et al. \textit{in prep}) and light curves for a set of stars complete down to TESS-band magnitude = 13.5 were extracted with the MIT Quick Look Pipeline (QLP, \citealp{Huang:QLP}). The QLP extracts light curves from the FFIs using a difference image analysis technique. After removing the effects of scattered earthshine and moonshine from the images, the QLP subtracts a high-quality reference image from each individual science frame and measures difference fluxes within sets of photometric apertures surrounding each star in the image. These difference fluxes are then converted to absolute brightness measurements by adding back the median flux expected from each star based on its \TESS-band magnitude. The QLP light curves have been used to discover dozens of planets \citep[e.g.,][]{Huang:2018, Rodriguez:2019, Huang:2020} and a few thousand planet candidates (N. Guerrero et al. \textit{submitted}). Additional information and a description of the QLP procedures are given by \citet{Huang:QLP}.  The QLP lightcurves are shown in Figure \ref{fig:fullLCs}. All five of these light curves were flattened using {\it Keplerspline}\footnote{\url{https://github.com/avanderburg/keplerspline}}, a spline fitting routine to divide out the best-fit stellar variability \citep{Vanderburg:2014}. The spacing of the spline break points for each system was determined by minimizing the Bayesian information criterion following the methodology from \citet{Shallue:2018}. After removing the stellar variability, we keep all data from one full transit duration before the transit until one full transit duration after the transit, and discard the remaining baseline data (which contains very little useful information but is computationally expensive to model).  We use these lightcurve segments for the global fitting of each system (see \S\ref{sec:GlobalModel}).

Because the QLP uses aperture photometry on difference frames using a median combined reference image, it does not always measure absolute transit depths, only difference fluxes. Transit depths measured by QLP, especially in crowded fields, might be dependent on the accuracy of the \TESS-band magnitudes from the \TESS\ Input Catalog \citep{Stassun:2018_TIC} to correct for contamination due to blending of nearby stars within the aperture. To check that the transit depths measured by QLP for these targets are reasonably accurate, we performed a spot check for TOI-1601 with light curves extracted using a more traditional simple aperture photometry method \citep{Vanderburg:2019}. We did not deblend the photometry in this test because TOI-1601 is in a relatively sparse field, as shown from our high resolution imaging (see \S\ref{sec:AO}). We ran an independent \texttt{EXOFASTv2} fit for TOI-1601 b (swapping this lightcurve for the QLP one) following the strategy discussed in \S\ref{sec:GlobalModel}, and the results were consistent to within $<$1$\sigma$ uncertainties. Additionally, for each system we used the follow-up ground-based photometry within the global analysis to also constrain the depth, providing independent constraints that can be used to confirm the QLP depths (since the TFOP photometry is at a higher angular resolution than \tess). Within our global fit, we checked on any unknown contamination by fitting for a dilution term on the \tess\ bandpass to account for any difference compared to the SG1 photometry. In all cases other than TOI-1478, the fitted dilution was consistent with zero and well within our Gaussian 10\% prior around zero, showing clear consistency between \tess\ and the TFOP seeing-limited photometry. TOI-1478 showed a significant required dilution on the order of 12\%. To properly account for this, we removed the \tess\ dilution prior (see \S\ref{sec:GlobalModel} for details), allowing it to be a free parameter, to properly correct for this within the fit. 

We searched the non-flattened QLP light curves for rotation-based modulations using the \texttt{VARTOOLS} Lomb-Scargle function \citep{Hartman:2016}. Specifically, we searched from 0.1 to 30 days and detect a clear strong periodicity at 5.296 days for TOI-1333, significantly different from the orbital period of its planetary companion (P$_b$ = 4.72 days). This same periodicity is observed in our ground-based photometry, ruling out any systematics in the \tess\ observations. We see some tentative evidence of a periodicity at 10-11 days for TOI-628, however it was only observed in one \tess\ sector. 

\begin{table*}
\scriptsize
\setlength{\tabcolsep}{2pt}
\centering
\caption{Literature and Measured Properties}
\begin{tabular}{llccccccccc}
  \hline
  \hline
Other identifiers\dotfill \\
& & TOI-628 & TOI-640 & TOI-1333 & TOI-1478& TOI-1601  \\
& & TIC 281408474& TIC 147977348 &TIC 395171208& TIC 409794137& TIC 139375960 \\
& & HD 288842& ---& BD+47 3521A& --- & --- \\
& TYCHO-2 & TYC 0146-01523-1 & TYC 7099-00846-1 &TYC 3595-01186-1 & TYC 5440-01407-1 & TYC 2836-00689-1\\
&2MASS &  J06370314+0146031 &  J06385630-3638462 & J21400351+4824243&  J08254410-1333356& J02332674+4100483 \\ 
&TESS Sector & 6 & 6,7& 15,16 & 7,8 & 18 \\
\hline
\hline
Parameter & Description & Value &Value &Value &Value &Value & Source\\
\hline 
$\alpha_{J2000}\ddagger$\dotfill&Right Ascension (RA)\dotfill & 06:37:03.13607&06:38:56.30742&21:40:03.50398&08:25:44.10708&02:33:26.74683&1\\
$\delta_{J2000}\ddagger$\dotfill&Declination (Dec)\dotfill & 01:46:03.19552&-36:38:46.14425&48:24:24.52541&-13:33:35.42756&+41:00:48.36893&1\\
&                \\
B$_T$\dotfill			&Tycho B$_T$ mag.\dotfill & 10.782$\pm$0.051&11.178$\pm$0.05&9.933$\pm$0.024&11.618$\pm$0.082&11.521$\pm$0.067&2\\
V$_T$\dotfill			&Tycho V$_T$ mag.\dotfill & 10.176$\pm$0.041&10.574$\pm$0.043&9.487$\pm$0.021&10.805$\pm$0.067&10.710$\pm$0.051&2\\
${\rm G}$\dotfill     & Gaia $G$ mag.\dotfill     &10.0579$\pm$0.02&10.4006$\pm$0.02&9.35$\pm$0.02&10.66$\pm$0.02&10.53$\pm$0.02&1\\
B$_{\rm P}$\dotfill			&Gaia B$_{\rm P}$ mag.\dotfill & 10.38$\pm$0.02&10.68$\pm$0.02&9.59$\pm$0.02&11.03$\pm$0.02&10.86$\pm$0.02&1\\
R$_{\rm P}$\dotfill			&Gaia R$_{\rm P}$ mag.\dotfill & 9.61$\pm$0.02&9.99$\pm$0.02&8.99$\pm$0.02&10.16$\pm$0.02&10.06$\pm$0.02&1\\
${\rm T}$\dotfill     & TESS mag.\dotfill     & 9.6565$\pm$0.0066& 10.0367$\pm$0.006&9.03527$\pm$0.0061&10.2042$\pm$0.006&10.1035$\pm$0.0065& 3 \\
&                \\
J\dotfill			& 2MASS J mag.\dotfill & 9.170$\pm$0.041&9.519$\pm$0.024&8.485$\pm$0.027& 9.590$\pm$0.023&9.505$\pm$0.022&4	\\
H\dotfill			& 2MASS H mag.\dotfill & 8.895$\pm$0.057&9.327$\pm$0.026&8.397$\pm$0.043&9.255$\pm$0.026&9.266$\pm$0.021 &4	\\
K$_S$\dotfill			& 2MASS ${\rm K_S}$ mag.\dotfill &8.811$\pm$0.02&9.243$\pm$0.023&8.272$\pm$0.024&9.201$\pm$0.021&9.19$\pm$0.02&4	\\
&                \\
\textit{WISE1}\dotfill		& \textit{WISE1} mag.\dotfill & 8.76$\pm$0.03 &9.213$\pm$0.03&7.706$\pm$0.013&9.15$\pm$0.03&9.16$\pm$0.03& 5	\\
\textit{WISE2}\dotfill		& \textit{WISE2} mag.\dotfill & 8.79$\pm$0.03 &9.240$\pm$0.03&7.594$\pm$0.012&9.19$\pm$0.03&9.21$\pm$0.03&  5	\\
\textit{WISE3}\dotfill		& \textit{WISE3} mag.\dotfill & 8.59$\pm$0.03 &9.239$\pm$0.03&8.035$\pm$0.019&9.18$\pm$0.03&9.18$\pm$0.034&   5	\\
\textit{WISE4}\dotfill		& \textit{WISE4} mag.\dotfill & 8.23$\pm$0.19 &---&8.043$\pm$0.163&---&8.80$\pm$0.33& 5	\\
&                \\
$\mu_{\alpha}$\dotfill		& Gaia DR2 proper motion\dotfill & -1.437$\pm$0.117&-3.872$\pm$0.040&-9.810$\pm$0.050&-8.277$\pm$0.058 &21.708$\pm$0.095&1 \\
                    & \hspace{3pt} in RA (mas yr$^{-1}$)	&&                \\
$\mu_{\delta}$\dotfill		& Gaia DR2 proper motion\dotfill 	&-1.082$\pm$0.087&4.927$\pm$0.045&-10.501$\pm$0.046&67.943$\pm$0.046&-0.874$\pm$0.090&1 \\
                    & \hspace{3pt} in DEC (mas yr$^{-1}$) &  &                \\
&                \\
$v\sin{i_\star}$\dotfill &  Rotational velocity (\kms) \hspace{9pt}\dotfill &  6.9$\pm$0.5&6.1 $\pm$ 0.5&  14.2$\pm$0.5& 4.3$\pm$0.5&6.4$\pm$0.5& \S\ref{sec:TRES}\& \S\ref{sec:TRES} \& \ref{sec:CHIRON}               \\
v$_{\rm mac}$\dotfill &  macroturbulent broadening (\kms) \hspace{9pt}\dotfill &  5.4$\pm$0.7&6.32$\pm$1.37&7.4$\pm$1.8& 4.9$\pm$0.5 & 6.3 $\pm$0.6& \S\ref{sec:TRES}\& \S\ref{sec:TRES} \& \ref{sec:CHIRON}      \\
$\pi^\dagger$\dotfill & Gaia Parallax (mas) \dotfill & 5.601$\pm$0.103 &2.925$\pm$0.033&4.989$\pm$0.038&6.542$\pm$0.047&2.974$\pm$0.080&1 \\
\\
$P_{\rm Rot}$\dotfill & Rotation Period (days)\dotfill &      &     &5.3$\pm$0.159&     &     & \S\ref{sec:TESS}, \ref{sec:wasp}, \& \ref{sec:kelt} \\
\hline
\hline
\end{tabular}
\begin{flushleft}
 \footnotesize{ \textbf{\textsc{NOTES:}}
 The uncertainties of the photometry have a systematic error floor applied. \\
 $\ddagger$ RA and Dec are in epoch J2000. The coordinates come from Vizier where the Gaia RA and Dec have been precessed and corrected to J2000 from epoch J2015.5.\\
 $\dagger$ Values have been corrected for the -0.30 $\mu$as offset as reported by \citet{Lindegren:2018}.\\
 $*$ $U$ is in the direction of the Galactic center. \\
 See \S D in the appendix of \citet{Collins:2017} for a description of each detrending parameter. 
 References are: $^1$\citet{Gaia:2018},$^2$\citet{Hog:2000},$^3$\citet{Stassun:2018_TIC},$^4$\citet{Cutri:2003}, $^5$\citet{Zacharias:2017}\\
}
\end{flushleft}
\label{tbl:LitProps}
\end{table*}

\subsection{WASP Photometry}
\label{sec:wasp}
The WASP transit search consisted of two, wide-field arrays of eight cameras, with SuperWASP on La Palma covering the northern sky and WASP-South in South Africa covering the south \citep{Pollacco:2006}.  Each camera used a 200-mm, f/1.8 lens with a broadband filter spanning 400--700 nm, backed by 2048x2048 CCDs giving a plate scale of $13.7\arcsec$ pixel$^{-1}$. Observations then rastered available fields with a typical 15-min cadence. 

We searched the WASP data for any rotational modulations using the methods from \citet{Maxted:2011}. TOI-640 was observed for spans of 150 nights in each of four years. The data from 2008, 2009 and 2010 show no significant modulation. The data from 2007, however, show significant power at a period of 62 $\pm$ 5 days, with an amplitude of 3 mmag and an estimated false-alarm probability below 1\%. Since this is seen in only one season, and given that the data span only 2.5 cycles, we do not regard this detection as fully reliable. For TOI-1333, the WASP data span 130 days in 2007, and show a clear modulation with a period of 15.9 $\pm$ 0.3 days, an amplitude of 19 mmag, and a false-alarm probability below 1\%. A similar periodicity is also seen in the \tess\ data and in the Kilodegree Extremely Little Telescope (KELT) data (see below) but at a third of this period at 5.3 days. No significant periodicity was detected for TOI-1478 or TOI-1601. 

\begin{table*}
 \centering
 \caption{Photometric follow-up observations of these systems used in the global fits and the detrending parameters.}
 \label{tbl:detrending_parameters}
 \begin{tabular}{llllllllll}
    \hline
    \hline
Target & Observatory & Date (UT) & size (m) & Filter & FOV & Pixel Scale  & Exp (s) & Additive Detrending\\
    \hline
TOI-628 b &FLWO/KeplerCam &2019 December 06 & 1.2 &$z^{\prime}$ & 23.1$\arcmin$ $\times$ 23.1$\arcmin$&  0.672$\arcsec$ & 9 & None \\
TOI-628 b &LCO TFN & 2019 December 16 & 0.4 & $z^{\prime}$ & 19$\arcmin$ $\times$ 29$\arcmin$ &   0.57$\arcsec$ & 18 & airmass \\
TOI-628 b &Whitin &2020 February 08& 0.7  &$r^{\prime}$ & 24$\arcmin$ $\times$ 24$\arcmin$    & 0.67$\arcsec$   &  10 &airmass\\
TOI-640 b & Brierfield & 2019 December 27 & 0.36 & $I$ &   49.4$\arcmin$ $\times$ 49.4$\arcmin$  &  1.47$\arcsec$ &   90  & X(FITS), Y(FITS)\\
TOI-640 b & PEST & 2020 March 01 & 0.3048 & $R$ & 31$\arcmin$ $\times$ 21$\arcmin$&  1.2$\arcsec$ & 60 & none \\
TOI-640 b &LCO SSO & 2020 August 23 & 1.0 & $z^{\prime}$ & 27$\arcmin$ $\times$ 27$\arcmin$ &   0.39$\arcsec$ & 60 & Y(FITS) \\
TOI-640 b &LCO SSO & 2020 November 06 & 1.0 & $z^{\prime}$ & 27$\arcmin$ $\times$ 27$\arcmin$ &   0.39$\arcsec$ & 60 & airmass \\
TOI-1333 b &CRCAO & 2020 July 29 & 0.6096 &$I$ & 26.8$\arcmin$ $\times$ 26.8$\arcmin$&  0.39$\arcsec$ & 45 & airmass \\
TOI-1333 b &LCO McDonald & 2020 July 29 & 0.4 & $z^{\prime}$ & 19$\arcmin$ $\times$ 29$\arcmin$ &   0.57$\arcsec$ & 30 & total counts \\
TOI-1333 b &LCO McDonald & 2020 August 12 & 0.4 & $z^{\prime}$ & 19$\arcmin$ $\times$ 29$\arcmin$ &   0.57$\arcsec$ & 30 & Y(FITS) \\
TOI-1333 b &CRCAO & 2020 September 19 & 0.6096 &$R$ & 26.8$\arcmin$ $\times$ 26.8$\arcmin$&  0.39$\arcsec$ & 80 & airmass \\
TOI-1478 b &FLWO/KeplerCam &2019 December 14 & 1.2 &$i^{\prime}$ & 23.1$\arcmin$ $\times$ 23.1$\arcmin$&  0.672$\arcsec$ & 7 & none \\
TOI-1478 b & PEST & 2020 January 03 & 0.3048 & $R$ & 31$\arcmin$ $\times$ 21$\arcmin$&  1.2$\arcsec$ & 30 & airmass \\
TOI-1601 b &GMU & 2020 August 30 & 0.8 &$R$ & 23$\arcmin$ $\times$ 23$\arcmin$ &   0.34$\arcsec$ & 30 & airmass, sky/pixels \\
TOI-1601 b &GMU & 2020 September 15 & 0.8 &$R$ & 23$\arcmin$ $\times$ 23$\arcmin$ &   0.34$\arcsec$ & 30 & airmass, sky/pixels, X(FITS) \\
TOI-1601 b &CRCAO & 2020 October 01 & 0.6096 &$z^{\prime}$ & 26.8$\arcmin$ $\times$ 26.8$\arcmin$&  0.39$\arcsec$ & 90 & airmass \\
TOI-1601 b & Adams   & 2020 October 17 & 0.61 &$I$ & 26$\arcmin$ $\times$ 26$\arcmin$ &   0.38$\arcsec$ & 60 & airmass, total counts \\
TOI-1601 b &LCO McDonald & 2020 October 17 & 1.0 &$z^{\prime}$ & 27$\arcmin$ $\times$ 27$\arcmin$ &   0.39$\arcsec$ & 60 & airmass, sky/pixels \\

        \hline
 \end{tabular}
\begin{flushleft}
{\footnotesize \textbf{\textsc{NOTES:}} All the follow-up photometry presented in this paper is available in machine-readable form in the online journal. 
}
\end{flushleft}
\end{table*}

Given knowledge of the \tess\ detections, transits of three of the systems described here can be found readily in the WASP data. TOI-640 was observed between 2006 and 2012, accumulating 23\,000 data points. The WASP search algorithm \citep{CollierCameron:2007} finds the transit with a period of 5.003773$\pm$0.000041 and a mid-transit epoch (T$_C$) of 2454822.00318$\pm$0.00411 HJD$_{\rm TDB}$. This detection had been overlooked by WASP vetters owing to the near-integer day period (5.00 days), since the dominant red noise in WASP data is at multiples of a day. TOI-1478 was observed between 2009 and 2012, accumulating 9000 data points, less than usual for WASP since the field is near the crowded Galactic plane. It had not been flagged as a WASP candidate, however the search algorithm finds the transit with a period of 10.18051$\pm$0.00017 days and a T$_C$ of 2455696.36710$\pm$0.00492 HJD$_{\rm TDB}$. TOI-1601 was observed over 2006 and 2007, accumulating 10,400 data points. The search algorithm finds the transit and gets a period of 5.33197$\pm$0.00010 and a T$_C$ of 2454186.65253$\pm$0.01283 HJD$_{\rm TDB}$. We use these T$_C$ values as priors for the \texttt{EXOFASTv2} global fits of TOI-640 b, TOI-1478 b, and TOI-1601 b. We see a $\sim$40\% reduction in uncertainty on the period of the planet when including the WASP T$_C$ prior. 

\subsection{KELT Photometry}
\label{sec:kelt}

To complement the \tess\ photometry, we analyzed observations of these five TOIs from the KELT survey\footnote{\url{https://keltsurvey.org}} \citep{Pepper:2007, Pepper:2012, Pepper:2018}. For a full description of the KELT observing strategy and reduction process, see \citet{Siverd:2012, Kuhn:2016}. KELT has two fully robotic telescopes, each of which uses a Mamiya 645 80mm f/1.9, each of which uses a Mamiya 645 80mm f/1.9 lens with 42mm aperture and Apogee 4k$\times$4k CCD on a Paramount ME mount. This setup provides to a $26\arcdeg\times26\arcdeg$ field of view with a $23\arcsec$ pixel scale. The two telescopes are located in at Winer Observatory in Sonoita, AZ and at the South African Astronomical Observatory (SAAO) in Sutherland, South Africa. KELT covers $\sim$85\% of the entire sky and the observing strategy results in a 20-30 minute cadence. Some of the KELT lightcurves are publicly available through the NASA Exoplanet Archive. KELT observations were available for TOI-628, TOI-1333, TOI-1478, and TOI-1601. KELT-South observed TOI-628 2828 times and TOI-1478 4632 times from 2010 to 2015. KELT-North observed TOI-1333 from 2012 to 2014 and TOI-1601 from 2006 to 2014, acquiring 2580 and 8520 observations, respectively. Since KELT has been observing since 2006 in some cases, the observations significantly extend the baseline of the photometry and can provide a strong constraint on the ephemeris of each system. Following the strategy described in \citet{Siverd:2012} and \citet{Kuhn:2016}, we also search the KELT lightcurves for transits of each planet. Unfortunately, no significant signs of the known planetary transits were found, likely due to the poor duty cycle for longer orbital periods.


Following the approach of \citet{Stassun:1999b} and \citet{Oelkers:2018}, we executed a search for periodic signals most likely to come from the rotation period of the star. For these stars, we post-processed the light curve data using the Trend-Filtering Algorithm \citep{Kovacs:2005} to remove common systematics. We then searched for candidate rotation signals using a modified version of the Lomb-Scargle period finder algorithm \citep{Lomb:1976, Scargle:1982}. We searched for periods between a minimum period of 0.5 days and a maximum period of 50 days using 2000 frequency steps\footnote{The total number of frequency steps may vary slightly depending on the number of data points in any given light curve.}. We masked periods between 0.5 and 0.505~days and 0.97--1.04~days to avoid the most common detector aliases associated with KELT's observational cadence and its interaction with the periods for the solar and sidereal day. For each star, we selected the highest statistically significant peak of the power spectrum (hereafter, $\gamma$) as the candidate period. 

We then executed a boot-strap analysis, using 1000 Monte-Carlo iterations, where the dates of the observations were not changed, but the magnitude values of the light curve were randomized, following the work of \citet{Henderson:2012}. We recalculated the Lomb-Scargle power spectrum for each iteration, and recorded the maximum peak power (hereafter, $\gamma_{sim}$) of all iterations. If the highest power spectrum peak was larger than the maximum simulated peak ($\gamma > \gamma_{sim}$) after 1000 iterations, we considered the periodic signal to be a candidate rotation period.

We find TOI-1333 to have a strongly significant ($\gamma > 50$) candidate rotation period at 5.3 days, which is consistent with \tess\ (see \S\ref{sec:TESS}), and TOI-1601 and TOI-1478 to have weakly significant ($\gamma > 10$) candidate rotation periods of 9.3 and 16.6 days, respectively. However, we do not see these periodicities in the \tess\ photometry, and they are likely aliases of the KELT observing strategy. KELT did not obtain observations of TOI-640.

\subsection{Ground-based Photometry from the {\it TESS} Follow-up Observing Program Working Group} \label{sec:sg1}
To refine the ephemerides and transit parameters for each system while ruling out false positive scenarios, we obtained photometric follow-up observations on all five systems from the {\it TESS} Follow-up Observing Program Working Group\footnote{\url{https://tess.mit.edu/followup/}}, sub-group 1 (SG1) for seeing-limited photometry. Specifically, the follow-up comes from the Las Cumbres Observatory (LCO) telescope network \citep{Brown:2013}, Whitin Observatory at Wellesley College, KeplerCam on the 1.2m telescope at Fred Lawrence Whipple Observatory (FLWO), Brierfield Observatory, PEST Observatory, C. R. Chambliss Astronomical Observatory (CRCAO) at Kutztown University, Adams Observatory at Austin College, and Suto Observatory. To schedule the photometric transit follow-up observations, we used the \texttt{TAPIR} software package \citep{Jensen:2013}. The data reduction and aperture photometry extraction was performed using \texttt{AstroImageJ} \citep{Collins:2017} for all follow-up data other than the PEST observations which were done using a custom software package \texttt{PEST Pipeline}\footnote{\url{http://pestobservatory.com/the-pest-pipeline/}}. The TFOP follow-up photometry for these five systems are shown in Figure \ref{fig:transits}, and a list of each telescope's information and details on each follow-up transit can be seen in Table \ref{tbl:detrending_parameters}. The follow-up transits presented in Figure \ref{fig:transits} are available as machine-readable tables with this paper.

\begin{deluxetable}{l l l l l}[bt]
\tabletypesize{\scriptsize}
\tablecaption{The first RV point from each spectrograph for all five systems. The full table of RVs for each system is available in machine-readable form in the online journal. \label{tab:rv}}
\tablewidth{0pt}
\tablehead{
\colhead{\bjdtdb} & \colhead{RV (m s$^{-1}$)} & \colhead{$\sigma_{RV}^{\dagger}$ (m s$^{-1}$)} & Target & \colhead{Instrument}
}
\startdata
2458604.64716 & 193.0 & 43.6 & TOI-628 & TRES \\
2458742.25270 & 19775.2 & 17.5 & TOI-628 & MINERVA3 \\
2458742.25270 & 19700.8 & 14.3 & TOI-628 & MINERVA4 \\
2458742.25270 & 19795.7 & 41.0 & TOI-628 & MINERVA6 \\
2458748.89482 & 39168.0 & 29.4 & TOI-640 & CHIRON \\
2458777.69583 & -50.6 & 48.0 & TOI-1333 & TRES \\
2458852.80172 & 19374.4 & 21.9 & TOI-1478 & CHIRON \\
2458875.60994 & 20738.5 & 22.0 & TOI-1478 & CORALIE \\
2458910.73537 & 20910.3 & 7.5 & TOI-1478 & FEROS \\
2458829.86525 & 67.4 & 27.6 & TOI-1478 & TRES \\
2459184.83629 & 19502.0 & 24.0 & TOI-1478 & CHIRON2\\
2458847.77575 & -184.1 & 32.0 & TOI-1601 & TRES\\
\hline
\enddata
\begin{flushleft}
{\footnotesize \textbf{\textsc{NOTES:}$^{\dagger}$ The internal RV error for the observation shown. }}
\end{flushleft}
\end{deluxetable}

\begin{figure*}
	\centering\vspace{.0in}
	\includegraphics[width=0.33\linewidth, trim={2.5cm 13.0cm 9.4cm 8.5cm}, clip]{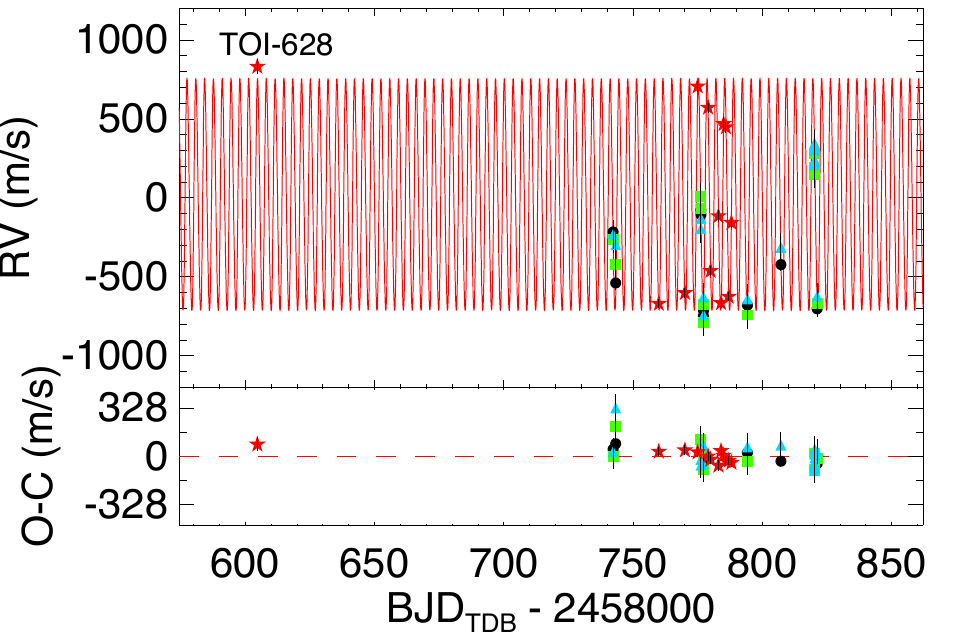}\includegraphics[width=0.33\linewidth, trim={2.5cm 13.0cm 9.4cm 8.5cm}, clip]{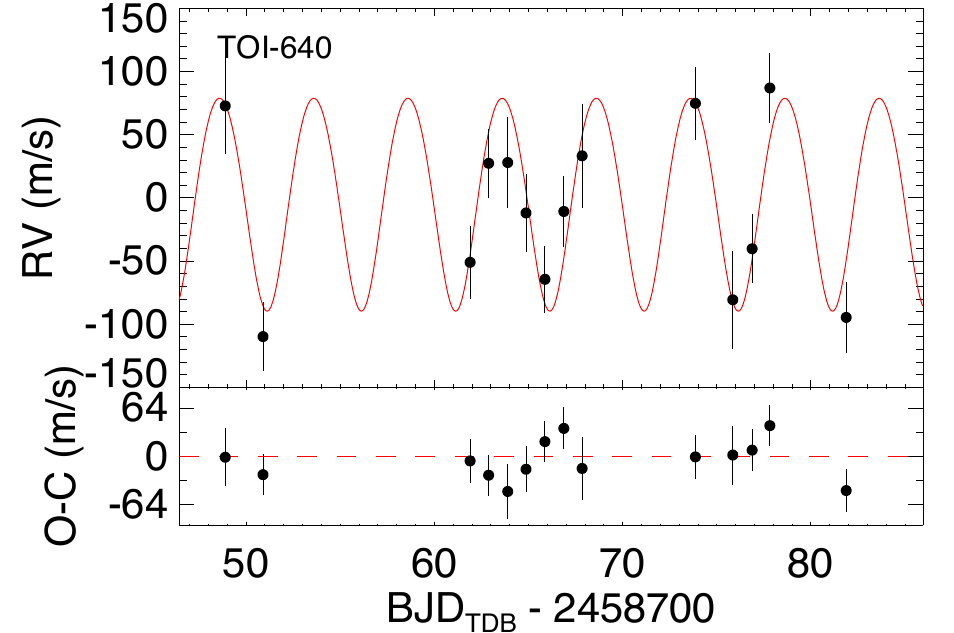}\includegraphics[width=0.33\linewidth, trim={2.5cm 13.0cm 9.4cm 8.5cm}, clip]{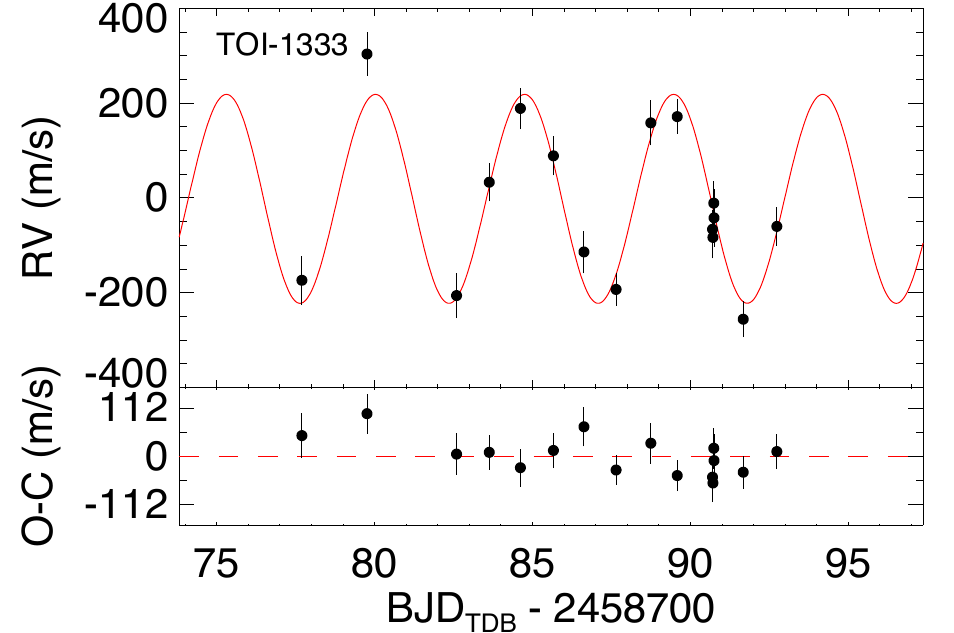}
	\includegraphics[width=0.33\linewidth, trim={2.5cm 13.0cm 9.4cm 8.5cm}, clip]{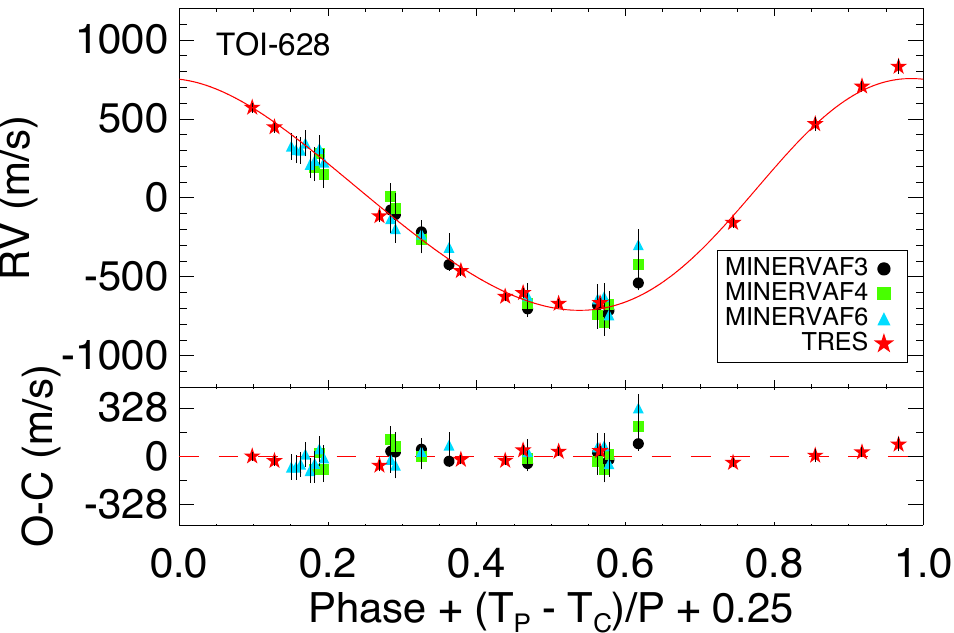}\includegraphics[width=0.33\linewidth, trim={2.5cm 13.0cm 9.4cm 8.5cm}, clip]{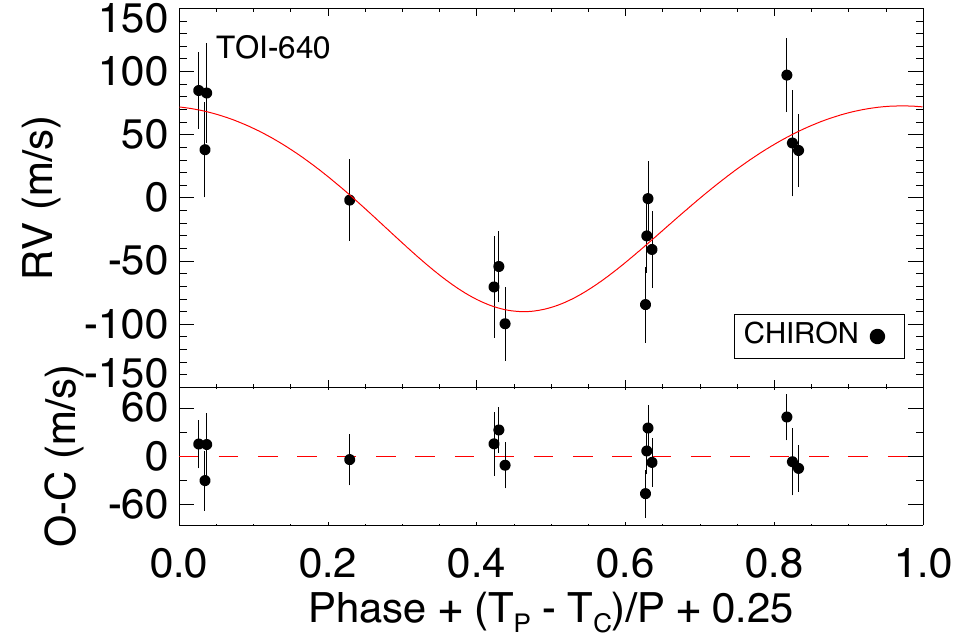}\includegraphics[width=0.33\linewidth, trim={2.5cm 13.0cm 9.4cm 8.5cm}, clip]{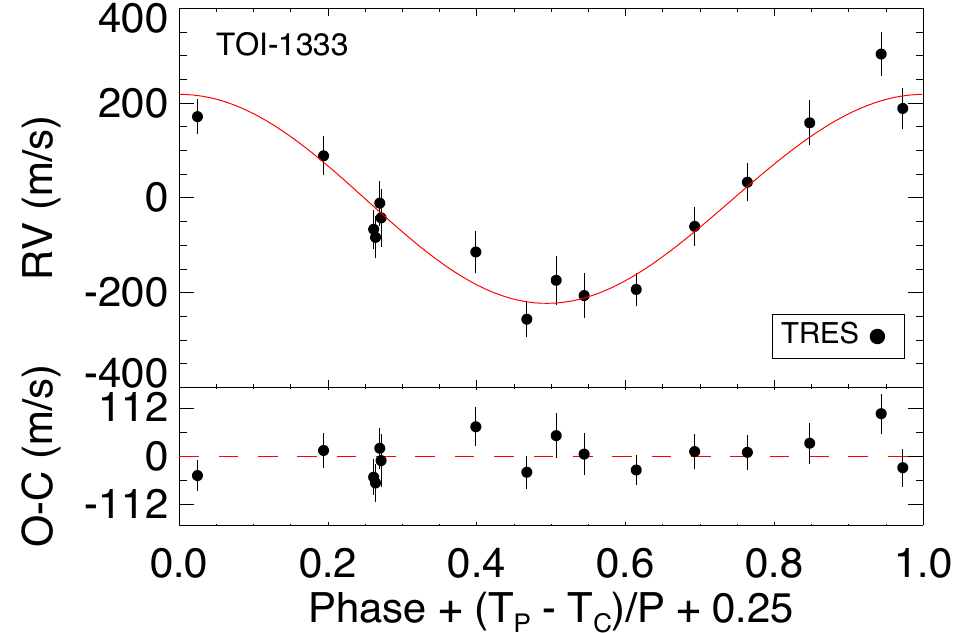}\\
	\includegraphics[width=0.33\linewidth, trim={2.5cm 13.0cm 9.4cm 8.5cm}, clip]{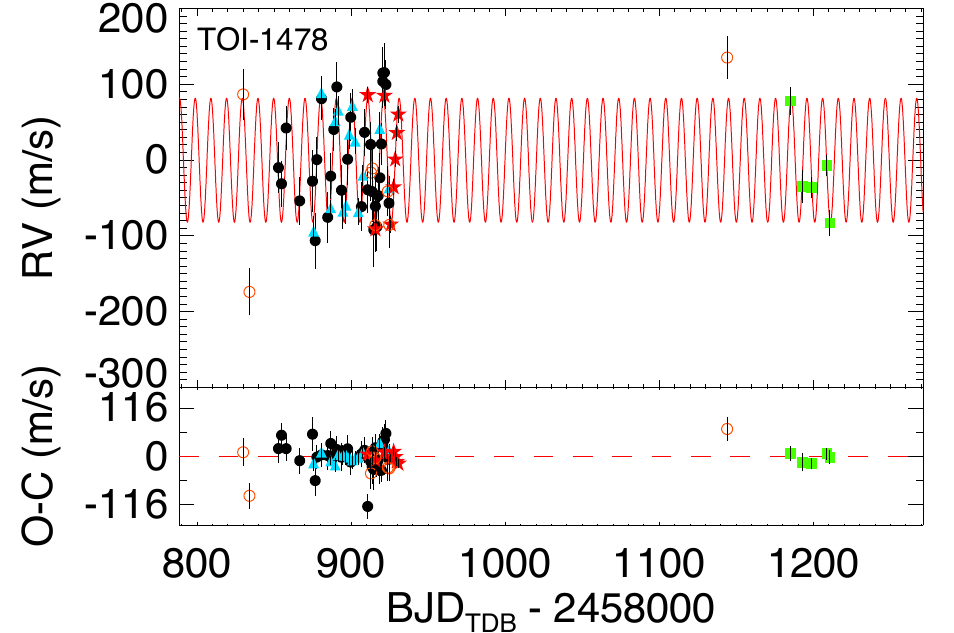}\includegraphics[width=0.33\linewidth, trim={2.5cm 13.0cm 9.4cm 8.5cm}, clip]{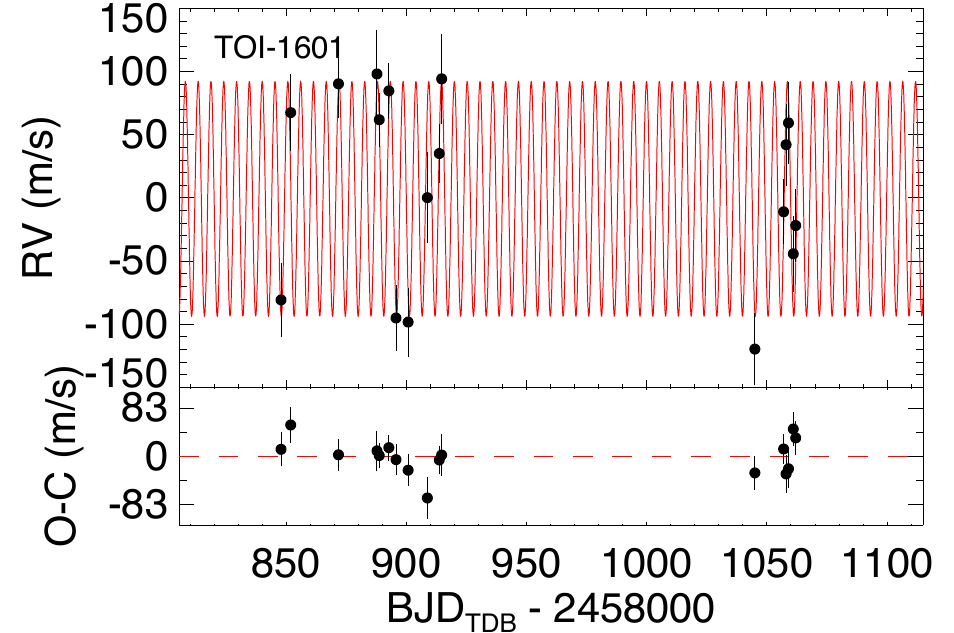}
	\includegraphics[width=0.33\linewidth, trim={2.5cm 13.0cm 9.4cm 8.5cm}, clip]{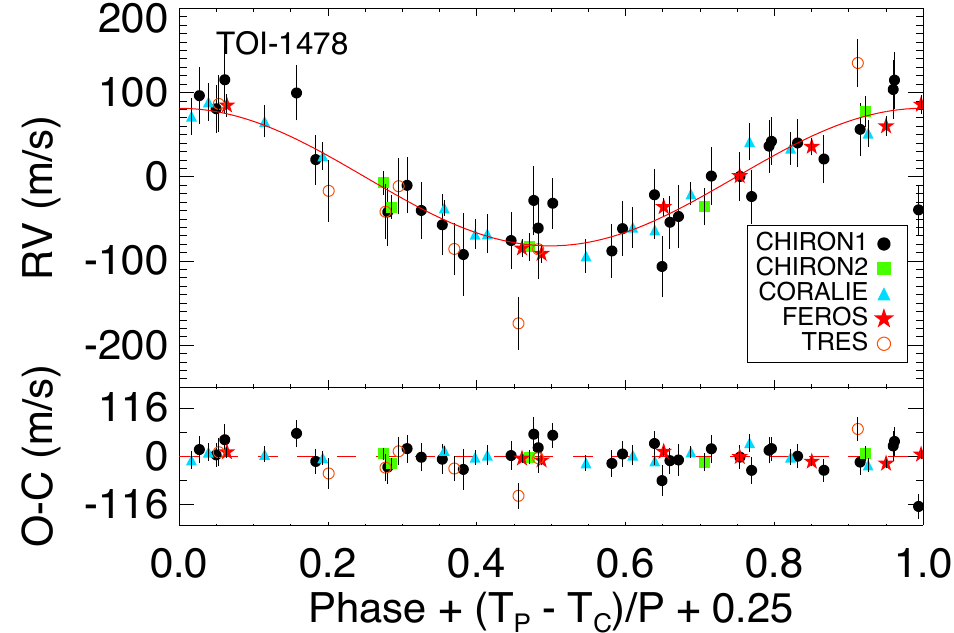}\includegraphics[width=0.33\linewidth, trim={2.5cm 13.0cm 9.4cm 8.5cm}, clip]{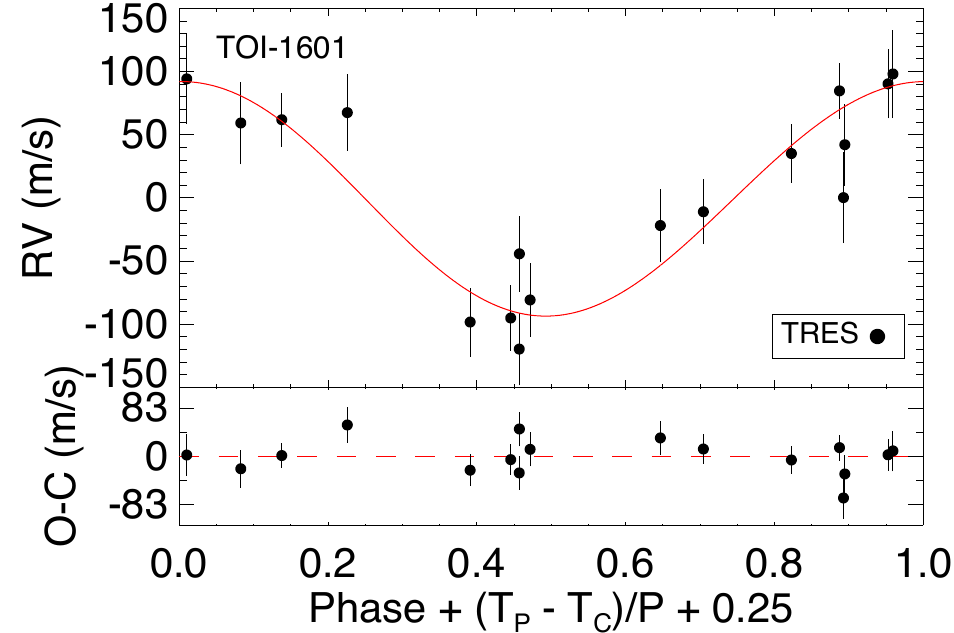}
	\caption{The RV observations of TOI-628 (top-left), TOI-640 (top-middle), TOI-1333 (top-right), TOI-1478 (bottom-left), and TOI-1601 (bottom-right). In each case, the top figure shows the RVs vs time and the bottom panel is phased to the best-fit ephemeris from our global fit. The \texttt{EXOFASTv2} model is shown in red and the residuals to the best-fit are shown below each plot. We see no periodicity in the residuals from our fit.}
	\label{fig:RVs} 
\end{figure*}

\begin{figure*}[!ht]
\includegraphics[width=0.33\linewidth,trim={1.5in 1in 0.5in 0.5in} ]{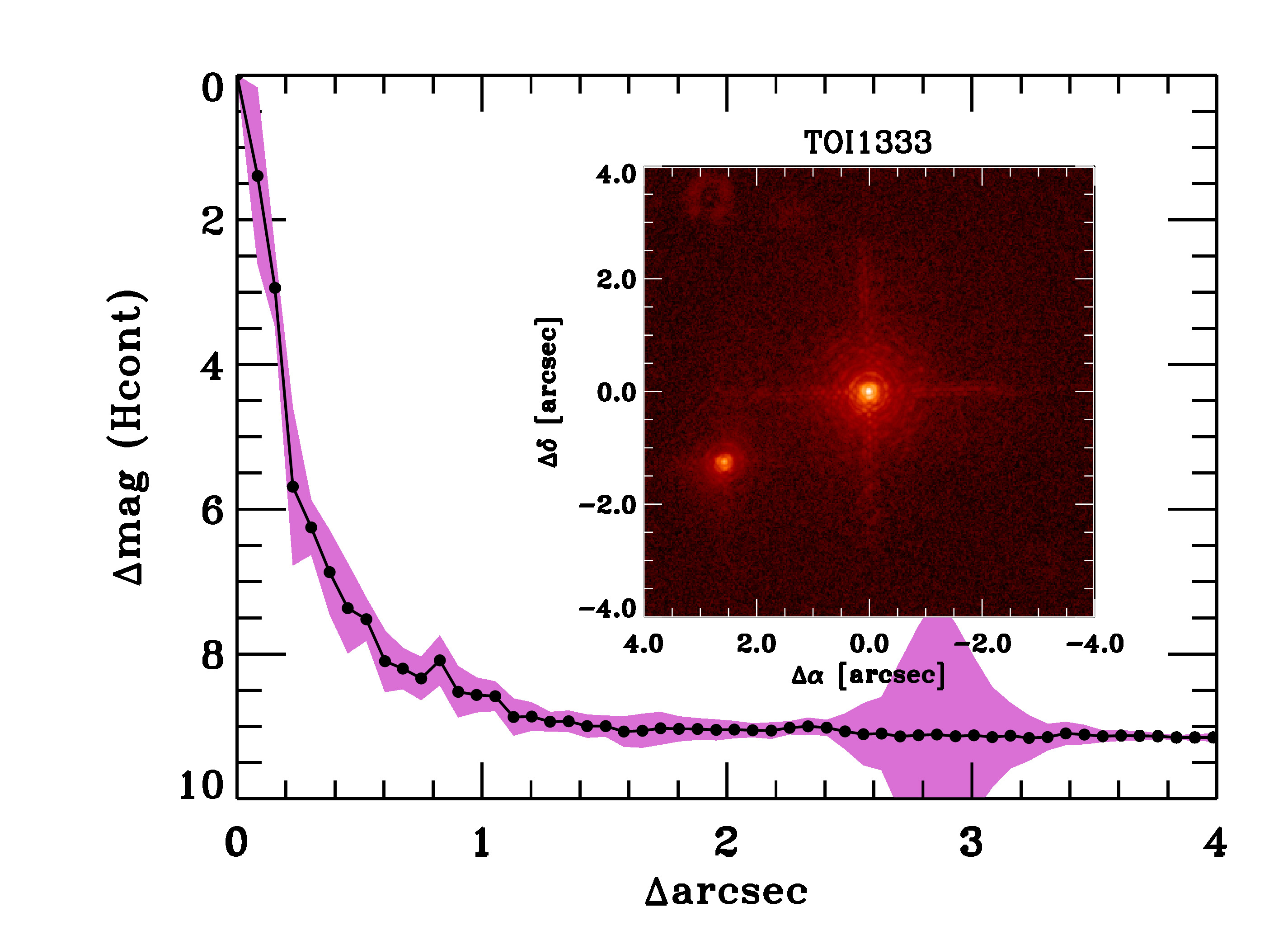}\includegraphics[width=0.33\linewidth,trim={1.5in 1in 0.5in 0.5in} ]{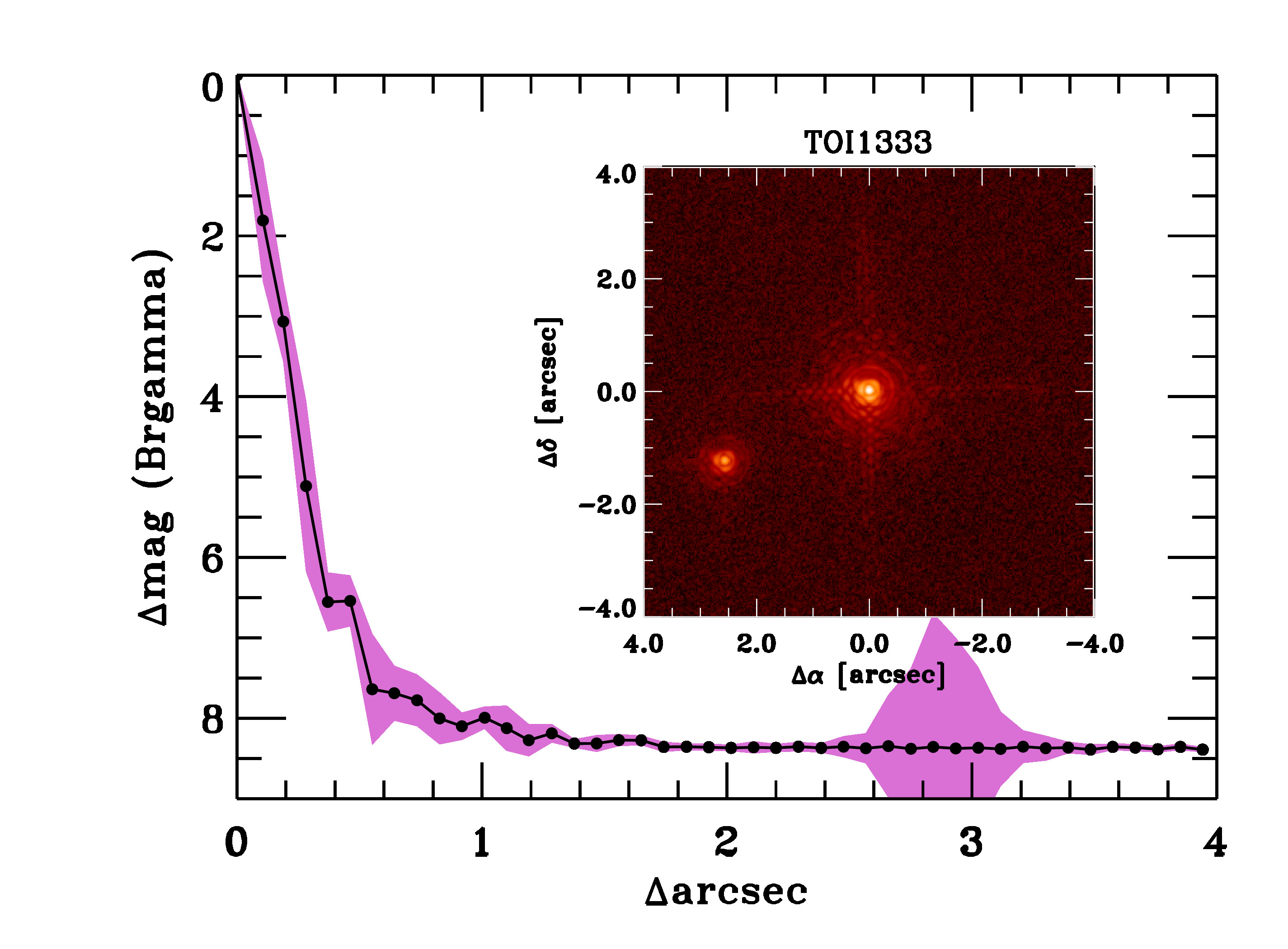}\includegraphics[width=0.32\linewidth,trim={0 0 0 0} ]{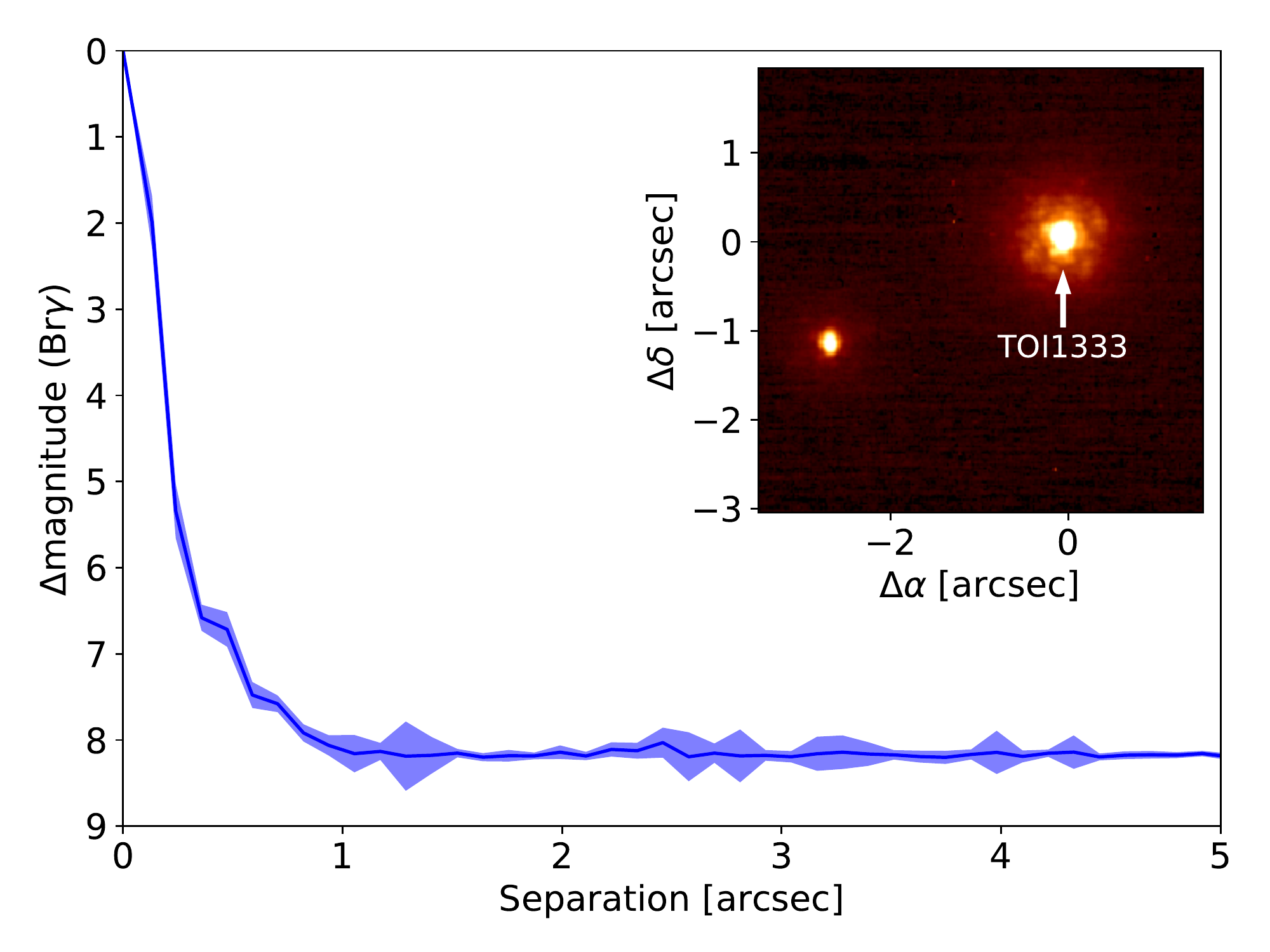}
\caption{The Palomar PHARO (left) H-band and (middle) Br$\gamma$-band 4$\sigma$ contrast curve for TOI-1333 with the AO image embedded in the plot. The (right) Gemini NIRI Br$\gamma$-band AO 5 $\sigma$ contrast curve for TOI-1333. The NIRI AO image is embedded in the plot. The second star in the image is TIC 2010985858, and we properly account for its blending in our fit (see \S\ref{sec:GlobalModel}). The colored swath represents the uncertainty on the 5$\sigma$ contrast curve (see \S\ref{sec:AO}).}
\label{fig:ao_fullfov}
\end{figure*}

\subsection{Spectroscopy}
\label{sec:spectroscopy}
To rule out false positive scenarios and measure the mass and orbital eccentricity of each system, we obtained time series spectroscopy coordinated through the TFOP WGs. A sample of one radial velocity (RV) point per target per instrument is shown in Table \ref{tab:rv}, with the full table available in machine-readable form in the online journal. The RVs and best-fit models from our \texttt{EXOFASTv2} analysis are shown in Figure \ref{fig:RVs} (see \S\ref{sec:GlobalModel}). Following the methodology in \citet{Zhou:2018}, we measure the $\vsini$ and macroturbulent broadening for all five systems from TRES except TOI-640 where the CHIRON observations were used (see Table \ref{tbl:LitProps}).

\subsubsection{TCES Spectroscopy} \label{sec:TLS}
Reconnaissance spectroscopic observations of TOI-628 were carried out with the Tautenburg Coud\'{e} Echelle Spectrograph (TCES) mounted at the 2-meter "Alfred Jensch" Telescope of the Thuringian State Observatory (TLS) in Tautenburg, Germany. The spectra cover the 470-740 nm wavelength range and have a resolution R=57000. A 40 min exposure was taken at BJD=2458777.6053 (orbital phase $\phi\sim$0.5), and a 3$\times$20 min exposure  at BJD=2458855.3720 ($\phi\sim$0.0). We measured for the two single-lined spectra a $\Delta RV \lesssim 1$ km s$^{-1}$, ruling out an eclipsing binary as the cause of the event detected by TESS. These velocities were not included in the global fit for TOI-628.

\subsubsection{TRES Spectroscopy} \label{sec:TRES}
To confirm targets from \tess\ in the Northern hemisphere, we observed TOI-628, TOI-1333, TOI-1478, and TOI-1601 with the Tillinghast Reflector Echelle Spectrograph \citep[TRES;][]{furesz:2008}\footnote{\url{http://www.sao.arizona.edu/html/FLWO/60/TRES/GABORthesis.pdf}} on the 1.5m Tillinghast Reflector. The telescope is located at the Fred L. Whipple Observatory (FLWO) on Mt. Hopkins, AZ, and the spectrograph has a resolving power of R=44,000. See  \citet{Buchhave:2010} and \citet{Quinn:2012} for a detailed description on reduction and RV extraction pipeline. The only difference in our analysis is that we created the template spectra for the RV extraction by aligning and median-combining all of the out-of-transit spectra. We removed cosmic rays and cross-correlated the median combined spectra against all the observed spectra. Bisector spans for the TRES spectra were calculated following the technique described in \citet{Torres:2007}. There was no correlation between the bisector spans and the RVs. We also used the TRES spectra to provide constraints on the \teff\ and \feh\ for our global analysis. We analyzed the TRES spectra with the Stellar Parameter Classification (SPC) package \citep{Buchhave:2012} to determine the metallicity and rotational velocity for all five host stars (see Table \ref{tbl:LitProps} and \ref{tab:exofast_stellar}). We also used SPC to determine a constraint on the \teff\ of 6250$\pm$100 K for TOI-1333, which is used in \S\ref{sec:sed} to constrain the dilution from nearby companions and the radius of TOI-1333. 


\subsubsection{CHIRON Spectroscopy} \label{sec:CHIRON}

We obtained a series of spectroscopic observations with the 1.5\,m SMARTS / CHIRON facility (see Table \ref{tab:rv} \citealp{Tokovinin:2013}) for TOI-640 and TOI-1478 to measure the host star parameters, and constrain their masses and eccentricities. The 1.5\,m SMARTS facility is located at Cerro Tololo Inter-American Observatory (CTIO), Chile. CHIRON is a high resolution echelle spectrograph fed via an image slicer through a single multi-mode fiber, with a spectral resolving power of $R\sim80{,}000$ over the wavelength region from 410 to 870nm. For the case of TOI-1478, we treat the pre- and post-shutdown RVs as separate instruments within the fit (see \S\ref{sec:GlobalModel}). 

To obtain the stellar atmospheric parameters, we matched the CHIRON spectra against an interpolated library of observed $\sim 10,000$ spectra classified by SPC \citep{Buchhave:2012}. The metallicity from this analysis was used as a prior for global fit of TOI-640 (see Table \ref{tab:exofast_stellar}). RVs were derived via the least-squares deconvolution \citep{Donati:1997, Zhou:2020} of the spectra against non-rotating synthetic templates matching the spectral parameters of each host star, generated using the ATLAS9 model atmospheres \citep{Kurucz:1992}. The RVs were then measured by fitting a least-squares deconvolution line profile with a rotational broadening kernel as prescribed by \citet{Gray:2005}. The velocities for each system are presented in Table \ref{tab:rv}.

\subsubsection{FEROS Spectroscopy} \label{sec:FEROS}
Using the FEROS spectrograph \citep{Kaufer:99} mounted on the 2.2m MPG telescope at La Silla observatory in Chile, we obtained 8 observations of TOI-1478. FEROS has R=48000, and the observations were between UT 2020 March 02 and UT 2020 March 22. A ThAr+Ne lamp was used to illuminate the fiber simultaneously to the observations to determine the instrumental offset. We reduced the spectra, derived the RVs, and produced the bisector spans using the \texttt{CERES} suite for echelle pipelines \citep{Brahm:2017}.

\subsubsection{CORALIE Spectroscopy} \label{sec:CORALUE}
TOI-1478 was observed with the CORALIE high resolution spectrograph (R=60,000) on the Swiss 1.2 m Euler telescope at La Silla Observatories, Chile \citep{Queloz:2001}. A total of 14 spectra were obtained between UT 2020 January 26 and March 16, covering several orbits of TOI-1478 b. CORALIE is fed by a 2\arcsec science fiber and a secondary fiber with simultaneous Fabry-Perot for wavelength calibration. RVs were computed with the standard CORALIE data reduction pipeline via cross-correlation with a binary G2 mask. Activity indices, bisector spans (BIS) and other line profile diagnostics were extracted as well. We find no correlation between the RVs and BIS, nor activity indicators. Our exposure times varied between 1200 and 1800 seconds depending on site conditions and observing schedule.

\subsubsection{MINERVA Australis Spectroscopy} \label{sec:MINERVA}
\textsc{Minerva}-Australis is an array of four PlaneWave CDK700 telescopes located in Queensland, Australia, fully dedicated to the precise RV follow-up of \tess\ candidates.  The four telescopes can be simultaneously fiber-fed to a single KiwiSpec R4-100 high-resolution (R=80,000) spectrograph \citep{barnes:2012, Addison:2019, Addison:2020}.  TOI-628 was monitored by \textsc{Minerva}-Australis using 2 or 3 telescopes in the array (Minerva3, Minerva4, Minerva6) between UT 2019 Sep 15 and Dec 3.  Each epoch consists of two 30-minute exposures, and the resulting RVs are binned to a single point as shown by the example in Table \ref{tab:rv}. Telescopes 1 and 3 obtained seven RV epochs, while Telescope 4 obtained four epochs. RVs for the observations are derived for each telescope by cross-correlation, where the template being matched is the mean spectrum of each telescope.  The instrumental variations are corrected by using the RVs computed from different Minerva telescopes as originating from independent instruments within our global model.

\subsection{High Resolution Imaging}
As part of our standard process for validating transiting exoplanets to assess possible contamination of bound or unbound companions on the derived planetary radii \citep{Ciardi:2015}, we obtained high spatial resolution imaging observations of all five systems. 

\subsubsection{Speckle Imaging}
\label{sec:SPECKLE}
We searched for close companions to TOI-628, TOI-640, and TOI-1478 with speckle imaging in the $I$-band on the 4.1-m Southern Astrophysical Research (SOAR) telescope \citep{Tokovinin:2018}. The speckle imaging was conducted using HRCam (field-of-view of 15$\arcsec$) and had a 0.01575$\arcsec$ pixel scale. TOI-628 was observed on UT 2019 November 11 with a sensitivity of $\Delta$Mag = 7.2 at 1$\arcsec$. Speckle observations of TOI-640 were taken on UT 2019 May 18 and had an estimated contrast of $\Delta$mag = 6.6 at 1$\arcsec$. Observations of TOI-1478 were taken on UT 2020 January 07 and had an estimated contrast of $\Delta$mag = 6.8 at 1$\arcsec$. See \citet{Ziegler:2020} for a description on the general observing strategy for \tess\ targets. No nearby companion was observed for any of the three targets out to 3$\arcsec$.

\begin{table*}
\tiny
\centering
\caption{Median values and 68\% confidence intervals for the global models}
\begin{tabular}{llccccccc}
  \hline
  \hline
  \\
\multicolumn{2}{l}{Priors:}& TOI-628 b &TOI-640 b &TOI-1333 b$^\mathsection$ & TOI-1478 b$^{\mathparagraph}$ &\multicolumn{2}{c}{TOI-1601 b}\\
\hline\\
Gaussian & $\pi$ Gaia Parallax (mas) \dotfill & 5.601$\pm$0.103 &2.925$\pm$0.033&---&6.542$\pm$0.047& \multicolumn{2}{c}{2.974$\pm$0.080}  \\
Gaussian & $[{\rm Fe/H}]$ Metallicity (dex) & 0.24$\pm$0.08 & 0.02$\pm$0.1 & 0.11$\pm$0.08 & 0.07$\pm$0.08 &  \multicolumn{2}{c}{0.31$\pm$0.08}\\
Upper Limit & $A_V$ V-band extinction (mag) & 2.977 & 0.292 & --- & 0.120 &  \multicolumn{2}{c}{0.14694}\\
Gaussian & $R_\star$ Stellar Radius (\rsun)& ---&--- &$1.963\pm0.064$& ---& --- & ---\\
Gaussian & \teff\ Stellar Effective Temperature (K)&--- &--- &$6250\pm100$&--- &--- &---\\
Gaussian & $T_C^{**}$ Time of conjunction (HJD$_{\rm TDB}$)\dotfill & ---&  2454822.00318$\pm$0.00411 & ---&2455696.36710$\pm$0.00492&\multicolumn{2}{c}{2454186.65253$\pm$0.01283}\\
Gaussian$^{\prime}$& $D_T$ Dilution in \tess\  \dotfill &0.00000$\pm$0.00344&0.00000$\pm$0.00360&0.00000$\pm$0.03817&---&\multicolumn{2}{c}{0.0$\pm$0.00196}\\
Gaussian$^{***}$& $D_I$ Dilution in $I$  \dotfill &---&---&0.3609$\pm$0.0180&---&\multicolumn{2}{c}{---}\\
Gaussian$^{***}$& $D_R$ Dilution in $R$  \dotfill &---&---&0.3499$\pm$0.0175&---&\multicolumn{2}{c}{---}\\
Gaussian$^{***}$& $D_{z^{\prime}}$ Dilution in $z^{\prime}$  \dotfill &---&---&0.0664$\pm$0.0033&---&\multicolumn{2}{c}{---}\\
  \hline
  \hline
Parameter & Units & Values & &  \\
\multicolumn{2}{l}{Stellar Parameters:}&&&&\\
Probability\dotfill & & 100 \%& 100 \%& 100 \%& 100\%& 68.4 \% & 31.6\% \\
~~~~$M_*$\dotfill &Mass (\msun)\dotfill & $1.311^{+0.066}_{-0.075}$& $1.536^{+0.069}_{-0.076}$& $1.464^{+0.076}_{-0.079}$& $0.947^{+0.059}_{-0.041}$& $1.517^{+0.053}_{-0.049}$& $1.340^{+0.042}_{-0.045}$\\
~~~~$R_*$\dotfill &Radius (\rsun)\dotfill & $1.345^{+0.046}_{-0.040}$& $2.082^{+0.064}_{-0.058}$& $1.925^{+0.064}_{-0.063}$& $1.048^{+0.030}_{-0.029}$& $2.186^{+0.074}_{-0.063}$& $2.185^{+0.086}_{-0.080}$\\
~~~~$L_*$\dotfill &Luminosity (\lsun)\dotfill & $2.48^{+0.41}_{-0.28}$& $6.83^{+0.41}_{-0.54}$& $5.17^{+0.49}_{-0.45}$&$0.971^{+0.036}_{-0.035}$& $5.40^{+0.35}_{-0.33}$& $5.25^{+0.35}_{-0.33}$\\
~~~~$F_{Bol}$\dotfill &Bolometric Flux$\times$10$^{-9}$ (cgs)\dotfill & $2.52^{+0.39}_{-0.27}$& $1.87^{+0.10}_{-0.14}$& ---& $1.32\pm0.45$&$1.505^{+0.52}_{-0.61}$& $1.485^{+0.59}_{-0.61}$\\
~~~~$\rho_*$\dotfill &Density (cgs)\dotfill & $0.762^{+0.067}_{-0.078}$& $0.240^{+0.024}_{-0.025}$& $0.289^{+0.031}_{-0.028}$& $1.16^{+0.14}_{-0.11}$& $0.205^{+0.016}_{-0.018}$& $0.181^{+0.020}_{-0.019}$\\
~~~$\log{g}$\dotfill &Surface gravity (cgs)\dotfill & $4.300^{+0.026}_{-0.036}$& $3.987^{+0.030}_{-0.036}$& $4.034^{+0.032}_{-0.033}$& $4.374^{+0.039}_{-0.032}$& $3.940^{+0.022}_{-0.025}$& $3.885^{+0.029}_{-0.031}$\\
~~~~$T_{\rm eff}$\dotfill &Effective Temperature (K)\dotfill & $6250^{+220}_{-190}$& $6460^{+130}_{-150}$& $6274\pm97$& $5597^{+83}_{-82}$& $5948^{+87}_{-89}$& $5910^{+96}_{-99}$\\
~~~~$[{\rm Fe/H}]$\dotfill &Metallicity (dex)\dotfill & $0.258^{+0.078}_{-0.080}$& $0.072^{+0.085}_{-0.076}$& $0.119^{+0.078}_{-0.077}$&$0.078^{+0.072}_{-0.066}$& $0.329^{+0.073}_{-0.075}$& $0.299\pm0.076$\\
~~~~$[{\rm Fe/H}]_{0}^\dagger$\dotfill &Initial Metallicity \dotfill & $0.271^{+0.070}_{-0.068}$& $0.174^{+0.088}_{-0.080}$& $0.212^{+0.083}_{-0.084}$& $0.110^{+0.069}_{-0.064}$& $0.347\pm0.069$& $0.296^{+0.070}_{-0.069}$\\
~~~$Age$\dotfill &Age (Gyr)\dotfill & $1.28^{+1.6}_{-0.91}$& $1.99^{+0.55}_{-0.40}$& $2.33^{+0.71}_{-0.56}$&$9.1^{+3.1}_{-3.9}$& $2.64^{+0.38}_{-0.39}$& $4.27^{+0.53}_{-0.42}$\\
~~~~$EEP^\ddagger$\dotfill &Equal Evolutionary Phase \dotfill & $331^{+31}_{-37}$& $380^{+19}_{-17}$& $381^{+20}_{-21}$& $400^{+16}_{-37}$& $403.1^{+7.9}_{-11}$& $452.9^{+4.3}_{-5.2}$\\
~~~~$A_V$\dotfill &V-band extinction (mag)\dotfill & $0.22^{+0.18}_{-0.14}$& $0.217^{+0.055}_{-0.095}$& ---& $0.062^{+0.039}_{-0.041}$& $0.103^{+0.032}_{-0.053}$& $0.086^{+0.043}_{-0.053}$\\
~~~~$\sigma_{SED}$\dotfill &SED photometry error scaling & $3.27^{+1.2}_{-0.73}$& $0.84^{+0.35}_{-0.22}$& ---&$0.70^{+0.29}_{-0.17}$& $1.17^{+0.42}_{-0.27}$& $1.20^{+0.44}_{-0.27}$\\
~~~~$\varpi$\dotfill &Parallax (mas)\dotfill & $5.64\pm0.10$& $2.925\pm0.033$& ---& $6.536\pm0.047$& $2.948^{+0.074}_{-0.075}$& $2.973^{+0.077}_{-0.076}$\\
~~~~$d$\dotfill &Distance (pc)\dotfill & $177.4^{+3.2}_{-3.1}$& $341.8^{+3.9}_{-3.8}$& ---& $153.0\pm1.1$& $339.2^{+8.8}_{-8.3}$& $336.4^{+8.9}_{-8.5}$\\
\multicolumn{2}{l}{Planetary Parameters:}&&&&\\
~~~~$P$\dotfill &Period (days)\dotfill & $3.4095675^{+0.0000070}_{-0.0000069}$& $5.0037775\pm0.0000048$& $4.720219\pm0.000011$& $10.180249\pm0.000015$& $5.331751\pm0.000011$& $5.331751\pm0.000011$\\
~~~~$R_P$\dotfill &Radius (\rj)\dotfill & $1.060^{+0.041}_{-0.034}$& $1.771^{+0.060}_{-0.056}$& $1.396^{+0.056}_{-0.054}$& $1.060^{+0.040}_{-0.039}$& $1.239^{+0.046}_{-0.039}$& $1.241^{+0.052}_{-0.049}$\\
~~~$M_P$\dotfill &Mass (\mj)\dotfill & $6.33^{+0.29}_{-0.31}$& $0.88\pm0.16$& $2.37\pm0.24$& $0.851^{+0.052}_{-0.047}$& $0.99\pm0.11$& $0.912^{+0.095}_{-0.10}$\\
~~~~$T_0^\star$\dotfill &Optimal conjunction Time (\bjdtdb)\dotfill & $2458629.47972\pm0.00039$& $2458459.73877^{+0.00071}_{-0.00083}$& $2458913.37033\pm0.00045$& $2458607.90338^{+0.00052}_{-0.00049}$& $2458990.55302^{+0.00081}_{-0.00079}$& $2458990.55306^{+0.00081}_{-0.00082}$\\
~~~~$a$\dotfill &Semi-major axis (AU)\dotfill & $0.04860^{+0.00080}_{-0.00094}$& $0.06608^{+0.00098}_{-0.0011}$& $0.0626^{+0.0011}_{-0.0012}$& $0.0903^{+0.0018}_{-0.0013}$& $0.06864^{+0.00079}_{-0.00075}$& $0.06586^{+0.00069}_{-0.00075}$\\
~~~~$i$\dotfill &Inclination (Degrees)\dotfill & $88.41^{+1.0}_{-0.93}$& $82.54^{+0.42}_{-0.59}$& $85.70^{+1.3}_{-0.65}$& $88.51^{+0.29}_{-0.22}$& $88.84^{+0.81}_{-1.1}$& $88.4^{+1.1}_{-1.3}$\\
~~~~$e$\dotfill &Eccentricity \dotfill & $0.072^{+0.021}_{-0.023}$& $0.050^{+0.054}_{-0.035}$& $0.073^{+0.092}_{-0.052}$& $0.024^{+0.032}_{-0.017}$& $0.036^{+0.044}_{-0.026}$& $0.057^{+0.050}_{-0.039}$\\
~~~~$\tau_{\rm circ}$\dotfill &Tidal circularization timescale (Gyr)\dotfill & $3.32^{+0.53}_{-0.61}$& $0.206^{+0.060}_{-0.057}$& $1.29^{+0.37}_{-0.33}$& $42.5^{+11}_{-8.1}$& $1.85\pm0.38$& $1.52^{+0.42}_{-0.38}$\\
~~~$\omega_*$\dotfill &Argument of Periastron (Degrees)\dotfill & $-74^{+13}_{-15}$& $159^{+86}_{-99}$& $105^{+64}_{-40}$& $-110^{+120}_{-130}$& $-180^{+110}_{-100}$& $120^{+57}_{-70}$\\
~~~~$T_{eq}$\dotfill &Equilibrium temperature (K)\dotfill & $1586^{+52}_{-40}$& $1749^{+26}_{-30}$& $1679^{+35}_{-34}$& $918\pm11$& $1619^{+24}_{-23}$& $1642^{+25}_{-24}$\\
~~~~$K$\dotfill &RV semi-amplitude (m/s)\dotfill & $713^{+20}_{-22}$& $78\pm14$& $223^{+22}_{-21}$& $82.5^{+4.0}_{-3.8}$& $87.3^{+9.2}_{-9.9}$& $87.6^{+8.9}_{-9.6}$\\
~~~~$R_P/R_*$\dotfill &Radius of planet in stellar radii \dotfill & $0.08108^{+0.00075}_{-0.00071}$& $0.08738^{+0.00091}_{-0.00086}$& $0.0745^{+0.0014}_{-0.0015}$&$0.1040\pm0.0015$& $0.05827^{+0.00071}_{-0.00070}$& $0.05834^{+0.00074}_{-0.00072}$\\
~~~~$a/R_*$\dotfill &Semi-major axis in stellar radii \dotfill& $7.78^{+0.22}_{-0.27}$& $6.82^{+0.22}_{-0.24}$& $6.98^{+0.24}_{-0.23}$& $18.54^{+0.70}_{-0.60}$& $6.76^{+0.17}_{-0.20}$& $6.48\pm0.23$\\
~~~~$Depth$\dotfill &Flux decrement at mid transit \dotfill & $0.00657\pm0.00012$& $0.00764^{+0.00016}_{-0.00015}$& $0.00555\pm0.00021$& $0.01083\pm0.00032$& $0.003395^{+0.000083}_{-0.000081}$& $0.003403^{+0.000087}_{-0.000083}$\\
~~~~$\tau$\dotfill &Ingress/egress transit duration (days)\dotfill & $0.01247^{+0.0012}_{-0.00061}$& $0.0454^{+0.0028}_{-0.0026}$& $0.0177^{+0.0027}_{-0.0033}$& $0.0208\pm-0.0022$& $0.01483^{+0.00084}_{-0.00034}$& $0.01503^{+0.0013}_{-0.00049}$\\
~~~~$T_{14}$\dotfill &Total transit duration (days)\dotfill & $0.1576^{+0.0015}_{-0.0013}$& $0.1502\pm0.0017$& $0.1934^{+0.0025}_{-0.0029}$& $0.1736\pm0.0023$& $0.2627\pm0.0020$& $0.2631^{+0.0022}_{-0.0021}$\\
~~~~$b$\dotfill &Transit Impact parameter \dotfill & $0.23^{+0.13}_{-0.15}$& $0.8763^{+0.0063}_{-0.0067}$& $0.504^{+0.085}_{-0.19}$& $0.481^{+0.050}_{-0.076}$& $0.136^{+0.13}_{-0.095}$& $0.18^{+0.14}_{-0.12}$\\
~~~~$T_{S,14}$\dotfill &Total eclipse duration (days)\dotfill & $0.1387^{+0.0063}_{-0.0055}$& $0.1481^{+0.0070}_{-0.019}$& $0.208^{+0.041}_{-0.013}$& $0.1736^{+0.0075}_{-0.0066}$& $0.263^{+0.014}_{-0.013}$& $0.279^{+0.024}_{-0.017}$\\
~~~~$\rho_P$\dotfill &Density (cgs)\dotfill & $6.58^{+0.70}_{-0.75}$& $0.195^{+0.042}_{-0.040}$& $1.08^{+0.18}_{-0.15}$& $0.88^{+0.13}_{-0.11}$& $0.640^{+0.097}_{-0.093}$& $0.589^{+0.098}_{-0.091}$\\
~~~~$logg_P$\dotfill &Surface gravity \dotfill &$4.145^{+0.031}_{-0.038}$&$2.841^{+0.078}_{-0.095}$& $3.479^{0.054}_{-0.056}$&$3.273^{+0.045}_{-0.043}$&$3.201^{+0.052}_{-0.059}$&$3.165^{+0.055}_{-0.061}$\\
~~~~$T_S$\dotfill &Time of eclipse (\bjdtdb)\dotfill & $2458467.566^{+0.031}_{-0.038}$& $2454824.46^{+0.12}_{-0.20}$& $2458717.445^{+0.097}_{-0.13}$& $2455701.431^{+0.092}_{-0.12}$& $2454189.289^{+0.095}_{-0.17}$& $2454189.27^{+0.12}_{-0.19}$\\
~~~~$e\cos{\omega_*}$\dotfill & \dotfill & $0.019^{+0.014}_{-0.018}$& $-0.012^{+0.038}_{-0.062}$& $-0.012^{+0.032}_{-0.042}$&$-0.002^{+0.014}_{-0.018}$& $-0.007^{+0.028}_{-0.049}$& $-0.013^{+0.035}_{-0.057}$\\
~~~~$e\sin{\omega_*}$\dotfill & \dotfill & $-0.068^{+0.024}_{-0.022}$& $0.007^{+0.042}_{-0.029}$& $0.053^{+0.10}_{-0.056}$&$0.000^{+0.030}_{-0.029}$& $0.001^{+0.028}_{-0.027}$& $0.031^{+0.044}_{-0.033}$\\
~~~~$d/R_*$\dotfill &Separation at mid transit \dotfill & $8.29^{+0.36}_{-0.40}$& $6.76^{+0.38}_{-0.51}$& $6.59^{+0.54}_{-0.77}$& $18.58^{+0.97}_{-0.88}$& $6.74^{+0.32}_{-0.36}$& $6.27^{+0.41}_{-0.47}$\\
  \hline
  \hline
\end{tabular}
\begin{flushleft}
  \footnotesize{
    \textbf{\textsc{NOTES:}}\\
See Table 3 in \citet{Eastman:2019} for a detailed description of all derived and fitted parameters.\\
$^\mathparagraph$ No TESS dilution prior was used for TOI-1478 b because initial fits showed a fitted dilution past the 10\% prior we used in the other systems. We fit for a dilution term within the fit for the \tess\ bandpass but with no prior.\\
$^\mathsection$ The SED was not included within the global fit for TOI-1333.\\
$^{**}$ T$_C$ prior comes from analysis of the WASP photometry (see \S\ref{sec:wasp}). We note that this time is in HJD$_{\rm TDB}$ while all data files and results here are $\bjdtdb$. The difference between these two time systems is on the order of seconds while the precision on T$_C$ used as a prior is on order of minutes, and therefore has no influence on the results. \\
$^{***}$Dilution prior for TOI-1333 comes from our 3-component SED analysis (see \S\ref{sec:sed}).\\
$^{\prime}$We assume the TESS correction for blending is much better than 10\%. We use a prior of 10\% of the determined blending from TICv8 \citep{Stassun:2018_TIC}.\\
$^\dagger$The initial metallicity is the metallicity of the star when it was formed.\\
$^\ddagger$The Equal Evolutionary Point corresponds to static points in a stars evolutionary history when using the MIST isochrones and can be a proxy for age. See \S2 in \citet{Dotter:2016} for a more detailed description of EEP.\\
$^\star$Optimal time of conjunction minimizes the covariance between $T_C$ and Period. This is the transit mid-point. \\
$^\pi$The tidal quality factor (Q$_S$) is assumed to be 10$^6$.\\ }
 \end{flushleft}
\label{tab:exofast_stellar}
\end{table*}

\begin{table*}
\centering
\tiny
\caption{Median values and 68\% confidence intervals for the global models}
\begin{tabular}{llcccccc}
  \hline
  \hline
TOI-628&&\\
\multicolumn{2}{l}{Wavelength Parameters:}&B&r'&z'&\TESS\\
~~~~$u_{1}$\dotfill &linear limb-darkening coeff \dotfill &$0.646^{+0.058}_{-0.061}$&$0.329\pm0.051$&$0.213^{+0.053}_{-0.052}$&$0.236^{+0.045}_{-0.046}$\\
~~~~$u_{2}$\dotfill &quadratic limb-darkening coeff \dotfill &$0.286\pm0.057$&$0.306\pm0.049$&$0.313\pm0.050$&$0.292^{+0.047}_{-0.048}$\\
~~~~$A_D$\dotfill &Dilution from neighboring stars \dotfill &--&--&--&$0.0012\pm0.0036$\\
\multicolumn{2}{l}{Telescope Parameters:}&MINERVAF3&MINERVAF4&MINERVAF6&TRES\\
~~~~$\gamma_{\rm rel}$\dotfill &Relative RV Offset (m/s)\dotfill &$19990^{+26}_{-28}$&$19969\pm35$&$20032^{+26}_{-25}$&$-633^{+20}_{-18}$\\
~~~~$\sigma_J$\dotfill &RV Jitter (m/s)\dotfill &$60^{+29}_{-18}$&$102^{+36}_{-24}$&$83^{+26}_{-20}$&$54^{+28}_{-20}$\\
~~~~$\sigma_J^2$\dotfill &RV Jitter Variance \dotfill &$3700^{+4400}_{-1800}$&$10400^{+8600}_{-4300}$&$7000^{+4900}_{-2900}$&$2900^{+3800}_{-1700}$\\
\multicolumn{2}{l}{Transit Parameters:}&\TESS&KeplerCam UT 2019-12-06 (z')&LCO SSO UT 2019-12-16 (B)& Whitin UT 2020-02-08 (r')\\
~~~~$\sigma^{2}$\dotfill &Added Variance \dotfill &$0.000000119^{+0.000000024}_{-0.000000020}$&$0.0000366^{+0.0000020}_{-0.0000018}$&$0.00001138^{+0.0000010}_{-0.00000091}$&$0.00000765^{+0.00000047}_{-0.00000044}$\\
~~~~$F_0$\dotfill &Baseline flux \dotfill &$0.999998^{+0.000039}_{-0.000038}$&$0.99958\pm0.00023$&$1.00071\pm0.00021$&$1.00125\pm0.00012$\\
~~~~$C_{0}$\dotfill &Additive detrending coeff \dotfill &---&---&$0.00806^{+0.00058}_{-0.00057}$&$-0.00277\pm0.00031$\\
\hline
  \hline
TOI-640&&\\
\multicolumn{2}{l}{Wavelength Parameters:}&I&z'&\TESS&V\smallskip\\
~~~~$u_{1}$\dotfill &linear limb-darkening coeff \dotfill &$0.191\pm0.051$&$0.170\pm0.036$&$0.216\pm0.046$&$0.376\pm0.050$\\
~~~~$u_{2}$\dotfill &quadratic limb-darkening coeff \dotfill &$0.311\pm0.049$&$0.313\pm0.035$&$0.314\pm0.046$&$0.329\pm0.049$\\
~~~~$A_D$\dotfill &Dilution from neighboring stars \dotfill &--&--&$0.0000\pm0.0022$&--\\
\multicolumn{2}{l}{Telescope Parameters:}&CHIRON\smallskip\\
~~~~$\gamma_{\rm rel}$\dotfill &Relative RV Offset (m/s)\dotfill &$39091^{+10.}_{-11}$\\
~~~~$\sigma_J$\dotfill &RV Jitter (m/s)\dotfill &$32.3^{+12}_{-8.3}$\\
~~~~$\sigma_J^2$\dotfill &RV Jitter Variance \dotfill &$1040^{+900}_{-470}$\\
\multicolumn{2}{l}{Transit Parameters:}&\TESS &Brierfield UT 2019-12-27 (I)&PEST UT 2020-03-01 (V)&LCO SSO UT 2020-08-23 (z')&LCO SSO UT 2020-11-06 (z')\smallskip\\
~~~~$\sigma^{2}$\dotfill &Added Variance \dotfill &$-0.0000000009^{+0.0000000093}_{-0.0000000081}$&$0.00000259^{+0.00000059}_{-0.00000051}$&$0.00000220^{+0.00000052}_{-0.00000045}$&$0.00000187^{+0.00000053}_{-0.00000041}$&$0.000000056^{+0.000000098}_{-0.000000080}$\\
~~~~$F_0$\dotfill &Baseline flux \dotfill &$1.000016\pm0.000022$&$0.99995\pm0.00018$&$0.99984\pm0.00019$&$0.99956\pm0.00023$&$1.00034\pm0.00012$\\
~~~~$C_{0}$\dotfill &Additive detrending coeff \dotfill &---&$-0.0183\pm0.0033$&--&$0.00080^{+0.00050}_{-0.00049}$&$-0.00052\pm0.00025$\\
~~~~$C_{1}$\dotfill &Additive detrending coeff \dotfill &---&$-0.0233\pm0.0033$&--&---&---\\
\hline
  \hline
TOI-1333&&\\
\multicolumn{2}{l}{Wavelength Parameters:}&I&R&z'&TESS\smallskip\\
~~~~$u_{1}$\dotfill &linear limb-darkening coeff \dotfill &$0.228\pm0.049$&$0.258\pm0.046$&$0.183\pm0.035$&$0.226^{+0.043}_{-0.044}$\\
~~~~$u_{2}$\dotfill &quadratic limb-darkening coeff \dotfill &$0.323\pm0.049$&$0.297\pm0.048$&$0.303\pm0.035$&$0.306\pm0.047$\\
~~~~$A_D$\dotfill &Dilution from neighboring stars \dotfill &$0.345\pm0.016$&$0.344\pm0.017$&$0.0671\pm0.0033$&$-0.016\pm0.028$\\
\multicolumn{2}{l}{Telescope Parameters:}&TRES\smallskip\\
~~~~$\gamma_{\rm rel}$\dotfill &Relative RV Offset (m/s)\dotfill &$122^{+15}_{-14}$\\
~~~~$\sigma_J$\dotfill &RV Jitter (m/s)\dotfill &$41^{+21}_{-18}$\\
~~~~$\sigma_J^2$\dotfill &RV Jitter Variance \dotfill &$1700^{+2100}_{-1200}$\\
\multicolumn{2}{l}{Transit Parameters:}&\TESS&CRCAO UT 2020-07-29 (I)&LCO McD UT 2020-07-29 (z')&LCO McD UT 2020-08-12 (z')&CRCAO UT 2020-09-19 (R)\smallskip\\
~~~~$\sigma^{2}$\dotfill &Added Variance \dotfill &$0.000000175^{+0.000000026}_{-0.000000023}$&$0.00000078^{+0.00000014}_{-0.00000012}$&$0.00000732^{+0.0000011}_{-0.00000097}$&$0.00000353^{+0.00000052}_{-0.00000046}$&$0.000000883^{+0.000000092}_{-0.000000084}$\\
~~~~$F_0$\dotfill &Baseline flux \dotfill &$0.999984\pm0.000036$&$0.99985\pm0.00012$&$0.99927^{+0.00024}_{-0.00025}$&$1.00245\pm0.00018$&$1.00015\pm0.00012$\\
~~~~$C_{0}$\dotfill &Additive detrending coeff \dotfill &---&$0.00033\pm0.00023$&$-0.00060\pm0.00065$&$0.01022\pm,0.00068$&$-0.00017^{+0.00021}_{-0.00022}$\\
\hline
\hline
TOI-1478&&\\
\multicolumn{2}{l}{Wavelength Parameters:}&R&i'&\TESS\\
~~~~$u_{1}$\dotfill &linear limb-darkening coeff \dotfill &$0.408\pm0.048$&$0.349\pm0.044$&$0.410\pm0.045$\\
~~~~$u_{2}$\dotfill &quadratic limb-darkening coeff \dotfill &$0.258\pm0.049$&$0.264\pm0.047$&$0.320^{+0.048}_{-0.047}$\\
~~~~$A_D$\dotfill &Dilution from neighboring stars \dotfill &--&--&$0.126^{+0.020}_{-0.021}$\\
\multicolumn{2}{l}{Telescope Parameters:}&CHIRON1&CHIRON2&CORALIE&FEROS&TRES\\
~~~~$\gamma_{\rm rel}$\dotfill &Relative RV Offset (m/s)\dotfill &$19384.9\pm7.0$&$19419\pm11$&$20833.5^{+5.2}_{-5.4}$&$20821.5\pm4.5$&$-30\pm12$\\
~~~~$\sigma_J$\dotfill &RV Jitter (m/s)\dotfill &$32.3^{+6.6}_{-5.5}$&$1.5^{+37}_{-1.5}$&$0.00^{+13}_{-0.00}$&$10.4^{+4.2}_{-4.5}$&$20.1^{+1.7}_{-3.3}$\\
~~~~$\sigma_J^2$\dotfill &RV Jitter Variance \dotfill &$1040^{+470}_{-320}$&$0^{+1500}_{-300}$&$-7^{+180}_{-91}$&$108^{+110}_{-74}$&$405^{+69}_{-120}$\\
\multicolumn{2}{l}{Transit Parameters:}&\TESS&KeplerCam UT 2019-12-14 (i')&PEST UT 2020-01-03 (R)\\
~~~~$\sigma^{2}$\dotfill &Added Variance \dotfill &$0.000000083^{+0.000000049}_{-0.000000039}$&$0.00000537^{+0.00000039}_{-0.00000036}$&$0.0000136^{+0.0000012}_{-0.0000011}$\\
~~~~$F_0$\dotfill &Baseline flux \dotfill &$1.000054\pm0.000061$&$1.00003\pm0.00016$&$1.00026\pm0.00022$\\
~~~~$C_{0}$\dotfill &Additive detrending coeff \dotfill &--&$0.00146\pm0.00035$&--\\
\hline
\hline
TOI-1601&&\\
\multicolumn{2}{l}{Wavelength Parameters:}&I&R&z'&\TESS\\
~~~~$u_{1}$\dotfill &linear limb-darkening coeff \dotfill &$0.281\pm0.050$&$0.285\pm0.050$&$0.215\pm0.047$&$0.305\pm0.045$\\
~~~~$u_{2}$\dotfill &quadratic limb-darkening coeff \dotfill &$0.305^{+0.050}_{-0.049}$&$0.260^{+0.049}_{-0.050}$&$0.281^{+0.049}_{-0.048}$&$0.297\pm0.048$\\
~~~~$A_D$\dotfill &Dilution from neighboring stars \dotfill &--&--&--&$-0.0005\pm0.0047$\\
\multicolumn{2}{l}{Telescope Parameters:}&TRES\\
~~~~$\gamma_{\rm rel}$\dotfill &Relative RV Offset (m/s)\dotfill &$-100.8^{+7.0}_{-7.3}$\\
~~~~$\sigma_J$\dotfill &RV Jitter (m/s)\dotfill &$0.00^{+21}_{-0.00}$\\
~~~~$\sigma_J^2$\dotfill &RV Jitter Variance \dotfill &$-110^{+540}_{-300}$\\
\multicolumn{2}{l}{Transit Parameters:}&\tess&GMU UT 2020-08-30 (R)&GMU UT 2020-09-15 (R)&CRCAO UT 2020-10-01 (R)&Adams UT 2020-10-17 (I)&LCO McD UT 2020-10-17 (z')\\
~~~~$\sigma^{2}$\dotfill &Added Variance \dotfill &$-0.000000044^{+0.000000024}_{-0.000000020}$&$0.00001324^{+0.0000010}_{-0.00000097}$&$0.0000105^{+0.0000020}_{-0.0000017}$&$0.00000134^{+0.00000025}_{-0.00000021}$&$0.00000066^{+0.00000017}_{-0.00000015}$&$0.00000069^{+0.00000016}_{-0.00000014}$\\
~~~~$F_0$\dotfill &Baseline flux \dotfill &$0.999969\pm0.000042$&$0.99318\pm0.00018$&$1.00037\pm0.00032$&$1.00043\pm0.00012$&$1.00036\pm0.00011$&$0.998577^{+0.000091}_{-0.000092}$\\
~~~~$C_{0}$\dotfill &Additive detrending coeff \dotfill &--&$-0.01148\pm0.00042$&$-0.015\pm0.010$&$-0.00039\pm0.00033$&$-0.00232\pm0.00034$&$-0.00039\pm0.00024$\\
~~~~$C_{1}$\dotfill &Additive detrending coeff \dotfill &--&$0.00226\pm0.00067$&$0.018\pm0.010$&--&$-0.00638\pm0.00063$&$-0.0015\pm0.0013$\\
~~~~$C_{2}$\dotfill &Additive detrending coeff \dotfill &--&--&$-0.00160\pm0.00053$&--&--&--\\
\hline
\hline
\label{tab:exofast_other}
\end{tabular}
\begin{flushleft}
  \end{flushleft}
\end{table*}

We also obtained two sets of high-resolution speckle images of TOI-1478. One was collected on UT 2020 January 14 using the Zorro instrument mounted on the 8-meter Gemini South telescope located on the summit of Cerro Pachon in Chile and the other on UT 2020 February 18 using the ‘Alopeke instrument mounted on the 8-m Gemini-North telescope located on the summit of Mauna Kea in Hawaii. These twin instruments simultaneously observe in two bands, i.e., $\frac{\lambda}{\Delta\lambda}$ = 832/40 nm and 562/54 nm, obtaining diffraction limited images with inner working angles 0.026$\arcsec$ and 0.017$\arcsec$, respectively. Each observation consisted of 6 minutes of total integration time at each telescope taken as sets of 1000$\times$0.06 second images. All the images were combined and subjected to Fourier analysis leading to the production of final data products including speckle reconstructed imagery (see \citealp{Howell:2011}). Both speckle imaging results showed similar contrast limits and revealed that TOI-1478 is a single star to contrast limits of 5 to 9 magnitudes (out to 1.17$\arcsec$, ruling out most main sequence companions to TOI-1478 within the spatial limits of $\sim$4 to 180 au (for d=153 pc, as determined from the Gaia DR2 parallax \citealp{Gaia:2018}).

\subsubsection{Adaptive Optics Imaging}
\label{sec:AO}
We observed TOI-628 on UT 2019 November 11 using the ShARCS adaptive optics system on the 3-m Shane Telescope at Lick Observatory. ShARCS has a field of view of 20$\arcsec$ $\times$ 20$\arcsec$ and a pixel scale of 0.033$\arcsec$ pixel$^{-1}$. We conducted our observations using a square 4-point dither pattern with a separation of 4$\arcsec$ between each dither position. Our observations were taken in natural guide star mode with high winds. We obtained one sequence of observations in Ks-band and a second sequence in J-band for a total integration time of 510 s in Ks-band and 225 s in J-band. See \citet{Savel:2020} for a detailed description of the observing strategy. Neither set of observations revealed any companions for TOI-628.

We observed TOI-1333 ($Br-\gamma$ and $H-cont$) and TOI-1478 with infrared high-resolution adaptive optics (AO) imaging at Palomar Observatory.  The Palomar Observatory observations were made with the PHARO instrument \citep{hayward:2001} behind the natural guide star AO system P3K \citep{dekany:2013}.  The observations were made on 2019~Nov~10 UT in a standard 5-point quincunx dither pattern with steps of 5\arcsec.  Each dither position was observed three times, offset in position from each other by 0.5\arcsec\ for a total of 15 frames. The camera was in the narrow-angle mode with a full field of view of $\sim25\arcsec$ and a pixel scale of approximately 0.025\arcsec\ per pixel.   Observations were made in the narrow-band Br-$\gamma$ filter $(\lambda_o = 2.1686; \Delta\lambda = 0.0326\mu$m) for TOI-1333 and TOI-1478, and in the $H-cont$ filter $(\lambda_o = 1.668; \Delta\lambda = 0.018\mu$m) for TOI-1333. The observations get down to a $\Delta$Mag = 6.54 (Br-$\gamma$) and = 7.52 ($H-cont$) for TOI-1333 and a $\Delta$Mag = 6.8 (Br-$\gamma$) for TOI-1478 (all at $\sim$0.5$\arcsec$).

TOI-1333 was also observed using NIRI on Gemini-North \citep{Hodapp:2003} on UT 2019 November 14 in the Br-$\gamma$ filter. NIRI has a 22$\arcsec\times22\arcsec$ field-of-view with a 0.022$\arcsec$ pixel scale. Our sequence consisted of nine images, each with exposure time 4.4s, and we dithered the telescope between each exposure. A sky background was constructed from the dithered frames, and subtracted from each science image. We also performed bad pixel removal and flatfielding, and then aligned and coadded frames. NIRI got down to a $\Delta$Mag = 6.7 at 0.472$\arcsec$.   

We also observed TOI-1601 using the Near Infrared Camera 2 (NIRC2) adaptive optics (AO) set up on the W. M. Keck Observatory in the Br-$\gamma$ filter and in the $J-cont$ filter on UT 2020 September 09. The NIRC2 detector has a 9.971 mas pixel$^{-1}$ using a 1024$\times$1024 CCD (field-of-view = 10$\arcsec\times10\arcsec$, \citealp{Service:2016}). Unfortunately, the lower left quadrant of the CCD is known to have higher than typical noise levels in comparison to the others. To avoid this issue, a 3-point dither pattern technique was used. The images were aligned and stacked after normal flat-field and sky background corrections. No nearby companions were seen down to a $\Delta$mag = 6.680 ($J-cont$) and = 6.402 (Br-$\gamma$) for TOI-1601 at 0.5$\arcsec$.


While the observing strategy differed, all of the AO data were processed and analyzed with a custom set of IDL tools.  The science frames were flat-fielded and sky-subtracted.  The flat fields were generated from a median combination of the dark subtracted flats taken on-sky.  The flats were normalized such that the median value of the flats is unity.  The sky frames were generated from the median average of the 15 dithered science frames; each science image was then sky-subtracted and flat-fielded.  The reduced science frames were combined into a single combined image using an intra-pixel interpolation that conserves flux, shifts the individual dithered frames by the appropriate fractional pixels, and median-coadds the frames (see Figure \ref{fig:ao_fullfov}).  The final resolution of the combined dither was determined from the full-width half-maximum of the point spread function; the resolutions of the $Br-\gamma$ and $H-cont$ images are 0.092\arcsec\ and 0.075\arcsec, respectively (Figure \ref{fig:ao_fullfov}).

The sensitivities of the final combined AO images were determined by injecting simulated sources azimuthally around the primary target every $20^\circ $ at separations of integer multiples of the central source's FWHM (\citealp{Furlan:2017},  Lund et al. {\it in preparation)}. The brightness of each injected source was scaled until standard aperture photometry detected it with $5\sigma $ significance. The resulting brightness of the injected sources relative to the target set the contrast limits at that injection location. The final $5\sigma $ limit at each separation was determined from the average of all of the determined limits at that separation and the uncertainty on the limit was set by the rms dispersion of the azimuthal slices at a given radial distance. The sensitivity curves for TOI-1333 are shown in Figure \ref{fig:ao_fullfov} along with an inset image zoomed to the primary target showing no other companion stars.

\subsubsection{TOI-1333 Companions}
In the case of TOI-1333, two additional sources were detected in the PHARO and GEMINI AO imaging (Figure ~\ref{fig:ao_fullfov}, only the 3$\arcsec$ companion is shown).  The first source is 7.43\arcsec\ \citep{Gaia:2018} to the east of the primary target and is separately resolved by the 2MASS survey (2MASS J21400422+4824221; TIC 395171213).  The second source is only 2.81\arcsec\ \citep{Gaia:2018} away and was not separately detected by 2MASS - although it was detected by Gaia and hence is in the \tess\ Input Catalog (TIC 2010985858). Based upon the differential magnitudes measured by Palomar, the deblended infrared magnitudes for the primary star and nearby companion are $K_1 = 8.355 \pm 0.024$ mag, $H_1 = 8.514 \pm 0.043$ mag ($H_1-K_1 = 0.159\pm 0.049$) and $K_2 = 10.614 \pm 0.026$ mag, $H_2 = 10.871 \pm 0.044$ mag ($H_2-K_2 = 0.257\pm 0.051$), respectively.  The 3\arcsec\ star has a Gaia magnitude of $G = 12.6221 \pm 0.0016$.

\begin{figure}[!ht]
\centering 
\includegraphics[trim = 1.3in 0.6in 1in 1.15in,width=\columnwidth]{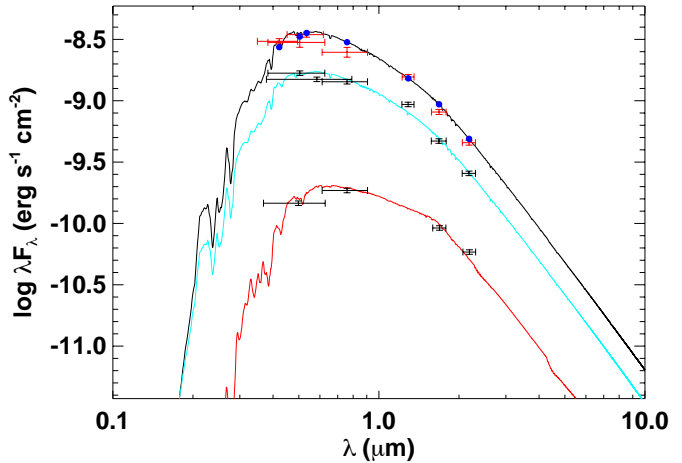}
\caption{The three-component SED fit for TOI-1333. The blue points are the predicted integrated fluxes for the primary star. The red and black points are the observed values at the corresponding passbands for each star. The width of the passbands are the horizontal error bars and the vertical errors represent the 1$\sigma$ uncertainties. The final model fit is shown by the solid line for TOI-1333 (black), and the 7$\arcsec$ (cyan) and bound 3$\arcsec$ (red) companions. }
\label{fig:sed_fit}
\end{figure}

The primary target has a Gaia distance of $200.5 \pm 1.2$pc whereas the 7\arcsec\ companion has a Gaia distance of $1030 \pm 35$ pc indicating that this companion is not bound to the primary and is simply a chance alignment near the line-of-sight to the TOI-1333. However, the 3\arcsec\ companion has a Gaia distance of $196.8 \pm 1.6$ pc and proper motions that are nearly identical to that of the primary star ($\mu1_{\alpha} = -9.81 \pm 0.05$ vs. $\mu2_{\alpha} = -9.06 \pm 0.08$ and $\mu1_{\delta} = -10.50 \pm 0.04$ vs. $\mu2_{\delta} = -9.24 ± 0.12$ mas/yr);  it is, therefore, highly probable that TIC 2010985858 is physically bound to the primary star with a projected separation of $\sim 470$ au. We account for the blended light from these two companions in our global analysis (see \S\ref{sec:GlobalModel} for details on how).

We use the LOFTI software package \citep{Pearce2020} to derive orbital parameters of the visual binary system of the formed by TOI 1333 and its companion. LOFTI uses the relative proper motions of the two stars from the Gaia catalog to sample probable orbits for a binary star system. To derive the mass of the companion -- required to fit orbit using LOFTI -- we use the \verb|isochrones| package \citep{Morton2015}. We perform an SED fit on photometry from TOI 1333's binary companion using the Gaia G, BP, and RP magnitudes, along with its Gaia parallax. Using the MIST isochrone \citep{Dotter:2016} as the base isochrone, we derive a mass for the companion of $0.808^{+0.043}_{-0.042}$ \msun. We use the astrometric parameters for the two systems from Gaia EDR3. 

The LOFTI fit reveals that the semimajor axis of the binary orbit is $570^{+590}_{-170}$ AU and the orbital inclination is $125^{+18}_{-10}$ degrees, ruling out an edge-on orbit for the binary at high confidence. The orbital eccentricity is weakly constrained to be less than 0.69 with 95\% confidence (with a slight preference for values between 0.5 and 0.7, but consistent with 0). 

\begin{figure*}[!ht]
	\centering\vspace{.0in}
	\includegraphics[width=0.95\linewidth, trim={1cm 5cm 1cm 5cm}, clip]{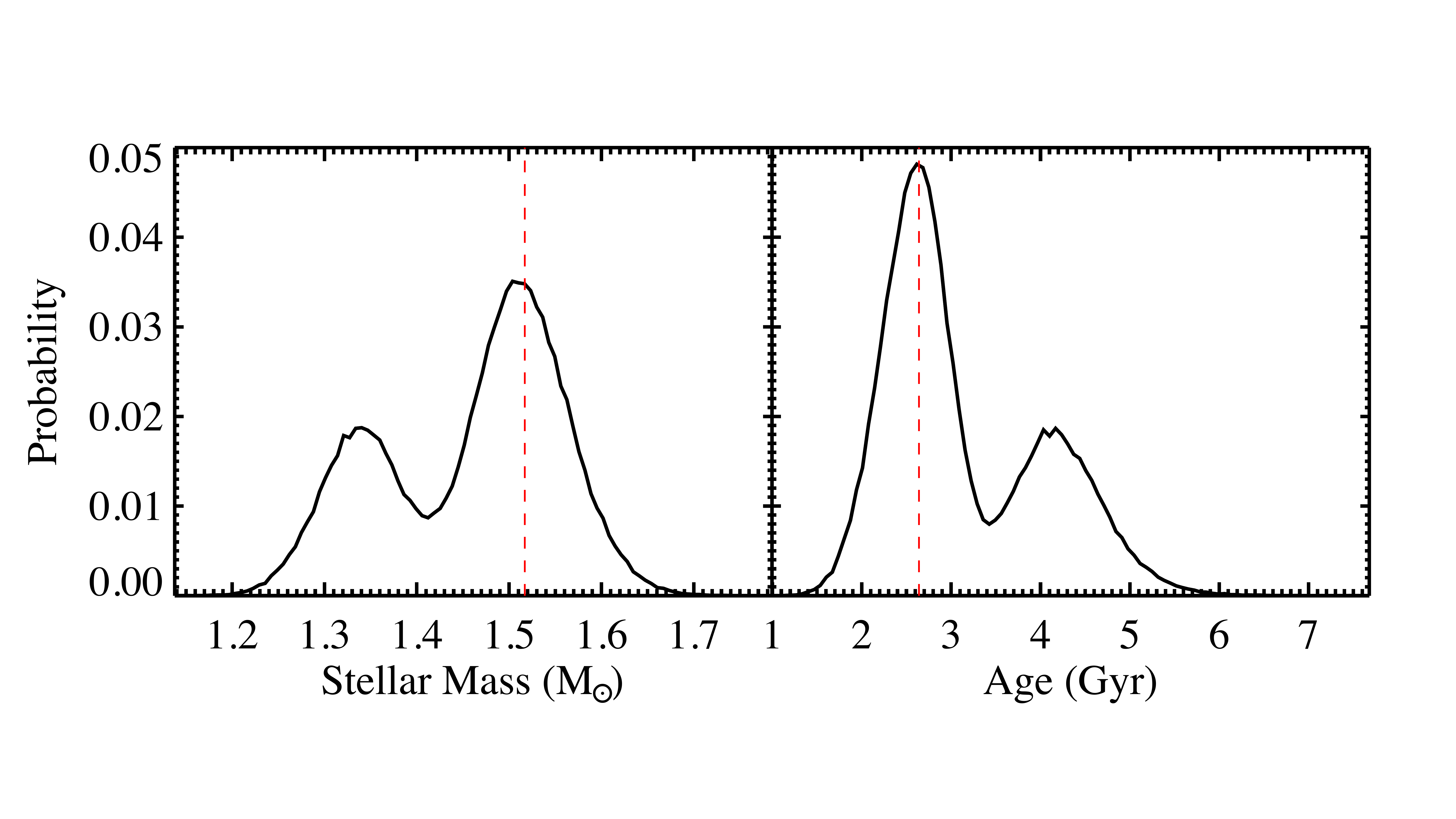}
	\caption{The (Left) M$_{\rm star}$ and (Right) Age probability distribution function for TOI-1601 from our global fit. We split this poster at the valley of \mstar\ = 1.415 \msun, and extract two separate solutions, one for each of the peaks in the posteriors (see Table \ref{tab:exofast_stellar}). The red line shows the median value for each parameter from the higher-mass solution with a probability of 68.4\% (see Section \ref{sec:GlobalModel}).}
	\label{fig:PDF} 
\end{figure*}

\subsubsection{TOI-1333 Spectral Energy Distribution}
\label{sec:sed}
The presence of the two stellar companions within a few arcseconds of TOI-1333 implies that the \tess\ and SG1 light curves of the TOI-1333 planet transit are likely to be diluted to some extent by the light from these other stars. Although the QLP corrects the \tess\ lightcurve for the blended contributions of known targets in the \tess\ input catalog (TIC), we need to correct the follow-up photometry from TFOP for different amounts of dilution. To quantify this flux dilution, we performed a multi-component SED fit with Kurucz model atmospheres following the procedures described in \citet{Stassun:2016}, utilizing the resolved broadband measurements from {\it Gaia}, {\it 2MASS}, and our AO observations (see \S\ref{sec:AO}). 

We adopted the spectroscopic \teff\ (6250$\pm$100 K) from TRES for TOI-1333 and the \teff\ from the TICv8 \citep{Stassun:2019} and from the {\it Gaia\/} DR2 catalog for the companions, with $A_V$ being left as a free parameter but limited to the maximum line-of-sight value from the dust maps of \citet{Schlegel:1998}. The resulting fits are shown in Figure~\ref{fig:sed_fit}, with best-fit $A_V = 0.06 \pm 0.06$ for the TOI-1333 planet host. Integrating the SED models over the \tess\ bandpass gives the total flux dilution $(F_2 + F_3)/F_1 = 0.55$ with values 0.56 in the $I$-band 0.54 in the $R$-band, and 0.58 in the Sloan z$^{\prime}$. We also used the SED fits to constrain the contribution from only the 2$\arcsec$ companion since some of our follow-up photometry only resolved the 7$\arcsec$ companion. The flux dilution of $F_2/F_1$ is 0.07 in the $I$-band, 0.06 in the $R$-band, and 0.07 in the Sloan z$^{\prime}$. By combining the deblended spectral energy distribution of the primary star TOI-1333 with the known Gaia DR2 parallax, we measure the its radius to be $R_\star = 1.963\pm0.064$~\rsun. We use this as a prior on the global fit for TOI-1333 (see \S\ref{sec:GlobalModel}).

\subsection{Location in the Galaxy, UVW Space Motion, and Galactic Population}
\label{sec:uvw}
For each of the TOIs analyzed here, we used their parallaxes, proper motions, and radial velocities and associated uncertainties from the Gaia DR2  catalog \citep{Gaia:2018} to determine their location, kinematics, orbits, and associations with known stellar populations\footnote{We acknowledge that some of the analysis in this section was inspired by and follows that of \citet{Burt:2020}}.  We corrected the native DR2 parallaxes and uncertainties using the prescription given in \citet{Lindegren:2018}.  From these we computed the {\it prior} (DR2-based) estimates of the distances to the systems\footnote{We note that, for self-consistency, we explicitly did not adopt the {\it posterior} values of the parallaxes from the global fit as listed in Table 4.}. We used these to compute the heliocentric UVW space motions of the host stars, and then corrected for the sun's motion (UVW)$_\odot$ with respect to the Local Standard of Rest (LSR) as determined by \citet{Coskunoglu:2011}.  These resulting (UVW) values are shown in Table 1.  We note that we adopt a right-handed coordinate system, such that positive $U$ is toward the Galactic center.

We used the Galactic latitudes and distances to the systems to estimate their $Z$ height relative to the sun, and then corrected for the $Z_\odot \simeq 30~{\rm pc}$ offset of the sun from the Galactic plane as determined by \citet{Bovy:2017} from the analysis of local giants.  We use the UVW velocities relative to the LSR to determine the likelihood that the star belongs to thin disk, thick disk, halo, or Hercules stream\footnote{The Hercules stream is a chemically heterogeneous population of nearby stars that have distinct kinematics relative to the bulk of the local stellar distribution.  See, e.g., \citealt{Bensby:2007}.}, using the categorization scheme of \citet{Bensby:2014}.  We also report the estimates of the parameters of the Galactic orbits of the systems as determined by \citet{Mackereth:2018} using Gaia DR2 astrometry and radial velocities\footnote{We note that \citet{Mackereth:2018} adopted a solar Galactocentric distance of $R_0=8~{\rm kpc}$, $Z_\odot =25~{\rm pc}$, and a local circular velocity of $V_{\rm circ}=220~{\rm km~s^{-1}}$. The also corrected for the sun's motion with respect to the LSR using the values of  (UVW)$_\odot$ as determined by  \citet{Schonrich:2010}, which differ slightly from the values we adopted as determined by \citet{Coskunoglu:2011}.  However, this is a minor effect and has no qualitative impact on our conclusions.}.  We estimated the spectral type of each TOI using its $T_{\rm eff}$ as determined from the global fit and given in Table 4, and using the $T_{\rm eff}$-spectral type relations of \citet{Pecaut:2013}.  We then compared the position and orbits of the systems to the scale height $h_Z$ of stars of similar spectral type as determined by \citet{Bovy:2017}.

Finally, we also consider whether the systems may belong to any of the known nearby ($\la 150~{\rm pc}$ young ($\la~{\rm Gyr}$) associations using the BANYAN $\Sigma$ (Bayesian Analysis for Nearby Young AssociatioNs $\Sigma$) estimator \citep{Gagne:2018}.  Not suprisingly, none of the systems had any significant ($\ga 1\%$) probability of being associated with these young associations, and the BANYAN $\Sigma$ estimator assigned all five systems as belonging to the `field' with a high probability $\ga 99\%$. We now discuss the results for each of the systems individually.

{\bf TOI-628:} We find a distance from the sun of $d=178\pm3~{\rm pc}$, consistent with the {\it posterior} value listed in Table 4, and $Z-Z_\odot \simeq -23~{\rm pc}$.  We derive velocities relative to the LSR of $(U,V,W)=(-8.7\pm 0.4,3.2\pm 0.2, 4.2\pm 0.1)~{\rm km~s^{-1}}$. According to the categorisation scheme of \citet{Bensby:2014}, the system has a $>99\%$ probability of belonging to the thin disk.  The Galactic orbit as estimated by \citet{Mackereth:2018} has a perigalacticon of $R_p=7.67~{\rm kpc}$, and apogalacticon of $R_a=8.19~{\rm kpc}$, an eccentricity of $e=0.03$, and a maximum $Z$ excursion from the Galactic plane of $Z_{\rm max}=63~{\rm pc}$.  This orbit is both consistent with the current location of the system, as well as the scale height of $97~{\rm pc}$ for stars of similar spectral type (F7V).  Indeed, TOI-628 is relatively dynamically `cold,' for its spectral type. In other words, it has an orbit that is fairly close to that of the local LSR.

{\bf TOI-640:} We find a distance from the sun of $d=340\pm4~{\rm pc}$, consistent with the {\it posterior} value listed in Table 4, and $Z-Z_\odot \simeq -76~{\rm pc}$.  We derive velocities relative to the LSR of $(U,V,W)=(-16.8\pm 0.2,-16.7\pm 0.4, -8.7\pm 0.2)~{\rm km~s^{-1}}$. According to the categorisation scheme of \citet{Bensby:2014}, the system has a $\sim 99\%$ probability of belonging to the thin disk.  The Galactic orbit as estimated by \citet{Mackereth:2018} has a perigalacticon of $R_p=6.28~{\rm kpc}$, and apogalacticon of $R_a=8.16~{\rm kpc}$, an eccentricity of $e=0.13$, and a maximum $Z$ excursion from the Galactic plane of $Z_{\rm max}=150~{\rm pc}$.  This orbit is both consistent with the current location of the system, and suggest that the system is nearing its maximum excursion above the plane.  It is also consistent as the scale height of $85~{\rm pc}$ for stars of similar spectral type (F5.5V). 

{\bf TOI-1333:} We find a distance from the sun of $d=200\pm2~{\rm pc}$, and $Z-Z_\odot \simeq 19~{\rm pc}$.  We derive velocities relative to the LSR of $(U,V,W)=(23.0\pm 0.1,-1.00\pm 0.3, -6.0\pm 0.1)~{\rm km~s^{-1}}$. According to the categorisation scheme of \citet{Bensby:2014}, the system has a $\sim 99\%$ probability of belonging to the thin disk.  The Galactic orbit as estimated by \citet{Mackereth:2018} has a perigalacticon of $R_p=7.25~{\rm kpc}$, and apogalacticon of $R_a=8.32~{\rm kpc}$, an eccentricity of $e=0.07$, and a maximum $Z$ excursion from the Galactic plane of $Z_{\rm max}=91~{\rm pc}$.  This orbit is consistent with the current location of the system.  It is also consistent as the scale height of $97~{\rm pc}$ for stars of similar spectral type (F7V).

{\bf TOI-1478:} We find a distance from the sun of $d=153\pm1~{\rm pc}$, and $Z-Z_\odot \simeq 67~{\rm pc}$.  We derive velocities relative to the LSR of $(U,V,W)=(-37.0\pm 0.3,26.4\pm 0.4, 32.5\pm 0.2)~{\rm km~s^{-1}}$. According to the categorisation scheme of \citet{Bensby:2014}, the system has a $\sim 88\%$ probability of belonging to the thin disk, and a $\sim 12\%$ probability of belonging to the thick disk (and negligible probabilities of belonging to the halo or Hercules stream).   The Galactic orbit as estimated by \citet{Mackereth:2018} has a perigalacticon of $R_p=7.71~{\rm kpc}$, and apogalacticon of $R_a=10.34~{\rm kpc}$, an eccentricity of $e=0.14$, and a maximum $Z$ excursion from the Galactic plane of $Z_{\rm max}=650~{\rm pc}$.  Unfortunately, \citet{Bovy:2017} was unable to determine the scale height of stars of similar spectral type (G6V) due to incompleteness.   Nevertheless, it would appear that TOI-1478's orbit has a maximum Z excursion that exceeds the expected scale height for stars of similar spectral type as estimated by extrapolating from the results of \citet{Bovy:2017} from earlier spectral types.  Surprisingly, its current distance above the plane is only small fraction of its predicted maximum excursion.  In summary, the weight of evidence suggests that TOI-1478 may well be a thick disk star that we happen to be observing when it is near the Galactic plane.  Detailed chemical abundance measurements (e.g., [$\alpha$/Fe]) may provide corroborating evidence for or against this hypothesis.

{\bf TOI-1601:} We find a distance from the sun of $d=336\pm9~{\rm pc}$, consistent with the {\it posterior} value listed in Table 4, and $Z-Z_\odot \simeq -73~{\rm pc}$.  We derive velocities relative to the LSR of $(U,V,W)=(-8.1\pm 0.7,-14.5\pm 0.7, 20.9\pm 0.4)~{\rm km~s^{-1}}$. According to the categorisation scheme of \citet{Bensby:2014}, the system has a $\sim 98\%$ probability of belonging to the thin disk.  The Galactic orbit as estimated by \citet{Mackereth:2018} has a perigalacticon of $R_p=6.50~{\rm kpc}$, and apogalacticon of $R_a=8.32~{\rm kpc}$, an eccentricity of $e=0.12$, and a maximum $Z$ excursion from the Galactic plane of $Z_{\rm max}=351~{\rm pc}$.  This orbit is consistent with the current location of the system.  The maximum $Z$ excursion is a factor of $\sim 3.3$ times larger than the scale height of $103~{\rm pc}$ for stars of similar spectral type (G0V). The probability that a star in a population with a given scale height $h_z=108~{\rm pc}$ has a maximum excursion of $z_{\rm max}=351$~{\rm pc} is non-negligible. Thus we expect that TOI-1601 is a thin disk star that is simply in the tail of the distribution of $z_{\rm max}$.  Again, detailed abundances could corroborate or refute this conclusion.

\section{EXOFAST\lowercase{v}2 Global Fits} 
\label{sec:GlobalModel}
We use the publicly available exoplanet fitting suite, \texttt{EXOFASTv2} \citep{Eastman:2013, Eastman:2019}, to globally fit the available photometry and RVs to determine the host star and planetary parameters for TOI-628 b, TOI-640 b, TOI-1333 b, TOI-1478 b, and TOI-1601 b. We fit the \tess\ and SG1 transits (see \S\ref{sec:sg1}), accounting for the 30 min smearing from the FFIs. Within the fit, the SG1 lightcurves were detrended (additive) against the corresponding parameters shown in Table \ref{tbl:LitProps}. See \S D in the appendix of \citet{Collins:2017} for a description of each detrending parameter. We use the MESA Isochrones and Stellar Tracks (MIST) stellar evolution models \citep{Paxton:2011, Paxton:2013, Paxton:2015, Choi:2016, Dotter:2016} and the spectral energy distribution (SED) within the fit to determine the host star parameters for all systems but TOI-1333 b. The SED fit within the global fit puts systematic floors on the broadband photometry errors (see Table \ref{tbl:LitProps}, \citealp{Stassun:2016}). We also note that EXOFASTv2 defaults a lower limit on the systematic error on the bolometric flux (F$_{\rm bol}\sim$3\%) given the spread seen from various techniques to calculate it \citep{Zinn:2019}. We place a Gaussian prior on the metallicity from our analysis of the host star's spectra from TRES, or CHIRON in the case of TOI-640 (see Section \ref{sec:CHIRON} and \ref{sec:TRES}). We also place a Gaussian prior on the parallax from Gaia \citep{Gaia:2016, Gaia:2018}, correcting for the 30 $\mu$as offset reported by \citet{Lindegren:2018}, and an upper limit on the line of sight extinction from \citet{Schlegel:1998} \& \citet{Schlafly:2011}. We also fit for a dilution term on the \tess\ band. Since the QLP corrects the \tess\ light curves for all known blended companions, we place a Gaussian prior of $0\pm$10\% of the contamination ratio reported by the \TESS\ Input Catalog (TIC,  \citealp{Stassun:2018_TIC}). We assume that the light curve has been corrected to a precision better than 10\% (and test this with preliminary \texttt{EXOFASTv2} showing the dilution to be consistent with zero), but this flexibility also provides an independent check on the correction applied and allows us to propagate the uncertainty in the correction. We do not find any significant additional dilution in any of the systems (within this prior and consistent with zero dilution) other than TOI-1478 b, where our fit suggests an additional 13\% dilution ($0.126^{+0.020}_{-0.021}$) is needed for the \tess\ light curve to be consistent with the TFOP photometry. For this fit, we remove this prior, essentially allowing the TFOP observations to constrain the depth of the transit. It is not clear what the cause of this additional dilution is since we see no evidence for any unknown companions in our high-spatial resolution imaging. We note that \TESS\ only observed three transits in one sector for TOI-1478 b, the longest period planet in our sample, and the TFOP light curves were both at higher spatial resolution. We also ran a fit where we allowed for a slope in the RVs, but found no significant trends for any system (we do not fit for a slope in the final fits). A list of the priors for each target is shown in Table \ref{tab:exofast_stellar}. Table 3 of \citet{Eastman:2019} shows a list and description of the fitted and derived parameters, including the bounds that EXOFASTv2 adopts for each fitted parameter. We note that eccentricity, a key parameter for this study, is bound as such: $0 \leq e \leq 1-\frac{a+R_p}{R_*}$, in order to ensure that the periastron values of the planet orbits are larger than the sum of the stellar and planetary radius. We deem a fit to be fully converged by following the recommending statistical threshold of a Gelman-Rubin statistic ($<$1.01) and independent draw ($>$1000) that is recommended by \citet{Eastman:2019}. The results from our \texttt{EXOFASTv2} fits are shown in Table \ref{tab:exofast_stellar} \& \ref{tab:exofast_other}, and the models are shown for the transits and RVs in Figures \ref{fig:transits} \& \ref{fig:RVs}.

In the case of TOI-1333 b, we deviate slightly from the methodology in the previous paragraph because there are two nearby bright companions, both detected by high resolution imaging (see \S \ref{fig:ao_fullfov}). The 2$\arcsec$ nearby companion and 7$\arcsec$ star were blended in the \tess\ and CRCAO photometry (see \S \ref{sec:sg1}), but only the 2\arcsec\ companion was blended in the LCO observations. While the \tess\ light curve has already been deblended as part of the reduction pipeline (see \S \ref{sec:TESS}), the SG1 observations were not. Our three-component SED analysis (see \S \ref{sec:sed}) determined that the nearby companion 2$\arcsec$ from TOI-1333 accounts for 6.6\% in the z$^{\prime}$-band, where LCO did not resolve the close companion. The combined flux contribution from both stars is 36.1\% in the $I$-band and 35.0\% in the $R$-band, where both companions were unresolved by CRCAO.  We use these values with a 5\% Gaussian prior \texttt{EXOFASTv2} global fit (also placing the prior on the \tess\ dilution as discussed in the previous paragraph). We place a Gaussian prior on the host star's radius from the SED analysis of $R_\star = 1.963\pm0.064$~\rsun. Preliminary SED fits of TOI-1333 using \texttt{EXOFASTv2} and independent \logg\ constraints from the SPC analysis of the TRES spectra suggested that TOI-1333 is a slightly evolved star. Given that the SED would normally constrain the \teff\ of the host star within the fit but was excluded for TOI-1333, we also place a prior on the \teff\ of 6250$\pm$100 K from the SPC analysis of the TRES spectroscopy. 

For TOI-1478 global fit, five RVs were acquired in the summer and winter of 2020, after a multi-month shutdown due to the COVID-19 pandemic. When included as part of the CHIRON RVs in the fit, we see a statistically significant slope measured of 0.176 m/s/day. However, the RV baseline for CHIRON observations has appeared to shift slightly when pre- and post-shutdown RVs were compared for standard stars, consistent with the shift we measured when including these post pandemic RVs (T. Henry, private communication). Since we saw no evidence of a slope in fits without these RVs and we know there is a shift observed in the RV baseline for CHIRON, we treat these observations as a separate instrument (labeled "CHIRON2"). When fitting the CHIRON RVs separately within the \texttt{EXOFASTv2} fit, we see a difference of 65 \ms\ in the fitted RV zero points. We note that the planet and host star parameters are consistent to $<$1 sigma whether or not this slope is included in the fit.

\subsection{TOI-1601 Bimodality}
\label{sec:TOI1601_bimodal}
After each \texttt{EXOFASTv2} fit, we inspect the posteriors of each fitted and derived parameter, visually inspecting for any anomalies such as multi-modal distributions. In all cases but TOI-1601 b, no issues were noted. For TOI-1601 b, we see a clear bimodal distribution in the mass and age of the host star (see Figure \ref{fig:PDF}). We find two peaks in the mass distribution at 1.340 and 1.517 \msun, which corresponds to the two peaks seen in the age distribution at 4.27 and 2.63 Gyr. There is no optimal way to represent the bimodal solution, so we split the mass of the host star at the minimum value of 1.415\msun\ and extract two solutions for both peaks identified. We adopt the high mass solution for the discussion since it has a higher probability (66.7\%) of being correct from our analysis, but present both solutions in Table \ref{tab:exofast_stellar} for future analysis. We note that we observed no significant change in any systematic parameters, suggesting the bimodality to be due to our limited precision that is not sufficient to completely separate similar solutions due to the host star being slightly evolved.

\begin{figure*}[ht]
	\centering\vspace{.0in}
	\includegraphics[width=0.99\linewidth, trim={0 0 0 0}, clip]{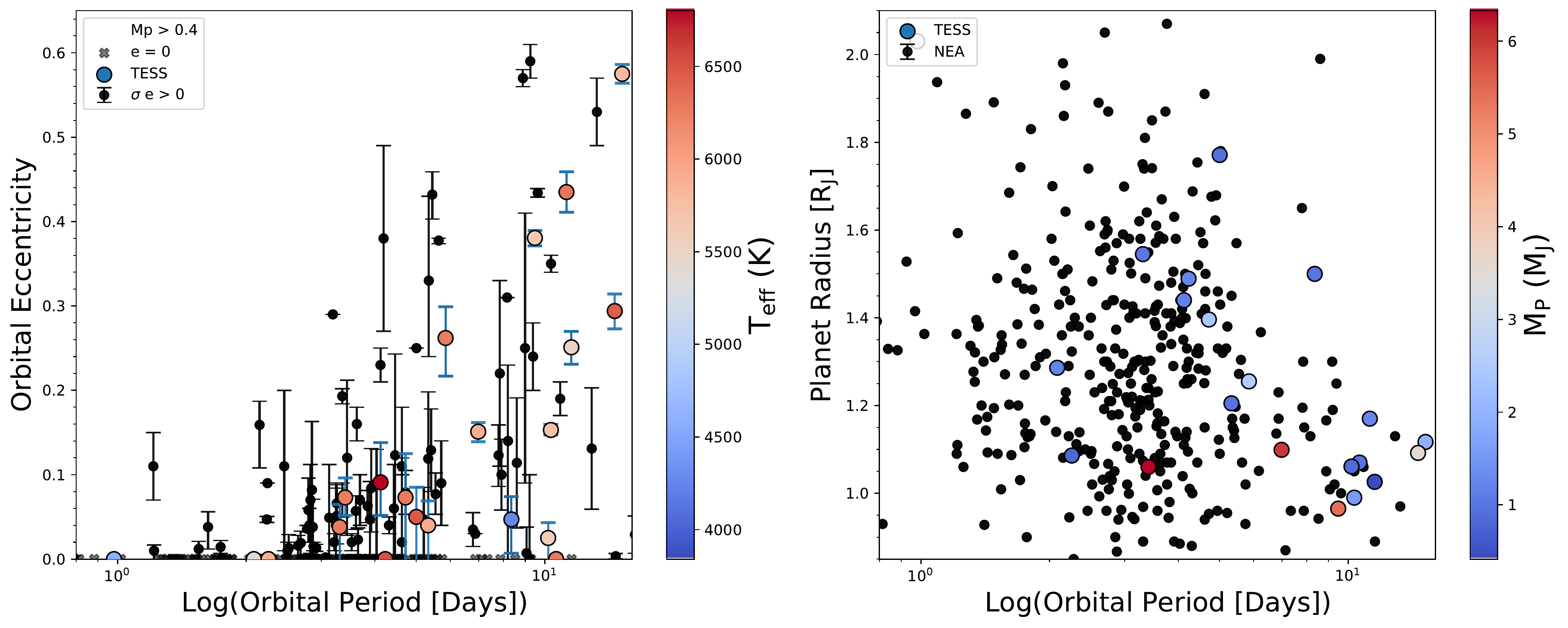}
\caption{(Left) The eccentricity and log of the orbital period of all known giant planets with a mass greater than 0.4\mj with period between 0.8 and 16 days. The \tess\ discovered systems are colored by the host star's effective temperature. The systems with a measured eccentricity from the NASA Exoplanet Archive (NEA) are shown as black circles with errors. Systems where the eccentricity was assumed to be zero are shown with gray crosses.  (Right) The radius and log of the orbital period of all known transiting giant planets. The systems known prior to \tess\ are in black, while the systems discovered by TESS, including those presented in this paper and \citet{IkwutUkwa:2021}, are shown as circles colored by their planet's mass.}
	\label{fig:discussion} 
\end{figure*}

\section{Discussion}
\label{sec:discussion}
These five newly-discovered hot Jupiters from NASA's \tess\ mission significantly increase ($>$5\%) the number of well characterized (measured masses and orbital eccentricities) giant planets that reside in orbits with periods between 5 and 15 days, a regime where planets experience weaker tidal forces than those experienced by planets closer to their host stars, and as a result, are not likely to have had enough time to circularize their orbits. Additionally, these longer period systems enable us to explore the "reinflation" scenario, an area \TESS\ should make a huge impact in given its high photometric precision. By continuing to discover and characterize new giant planets at longer periods, we can look for evidence (through their eccentricity distribution) of the dominant migration mechanism. Additionally, \tess\ will provide a complete magnitude-limited sample of hot Jupiters (P$<$10 days, \citealp{Zhou:2019}), allowing us to test whether multiple populations exist within the  distribution of key parameters (mass, radius, eccentricity), where some tentative trends have been suggested \citep{Nelson:2017, IkwutUkwa:2021}. Here we provide a short overview of our global fit results on each of the five new systems. We note that all three of the planets found to be orbiting subgiant host stars (TOI-640, TOI-1333, and TOI-1601) reside in circular orbits. 

\subsection{TOI-628 b}
TOI-628 (TIC 281408474) is a $V$=10.18 late-F star with a mass of $M_{*}$ = $1.311^{+0.066}_{-0.075}$ $M_{\odot}$, radius of $R_{*}$ = $1.345^{+0.046}_{-0.040}$ $R_{\odot}$, and an age of $1.28^{+1.6}_{-0.91}$ Gyr. Its planetary companion (TOI-628 b) has a radius of $R_{\rm P}$ = $1.060^{+0.041}_{-0.034}$ $R_{\rm J}$, a mass of $M_{\rm P}$ = $6.33^{+0.29}_{-0.31}$ $M_{\rm J}$, and is on a 3.4096-day period orbit. Our global analysis measures a non-zero orbital eccentricity of e = $0.072^{+0.021}_{-0.023}$. Within our global fit, we derive a circularization timescale of $\tau_{\rm circ}$ = $3.32^{+0.53}_{-0.61}$ Gyr (for this system, assuming equilibrium tides and tidal quality factors for the planet and star of $Q_p~10^6$ and $Q_*~10^6$ \citealp{Adams:2006}) for this system, which is longer than our estimated age from MIST of $1.28^{+1.6}_{-0.91}$ Gyr. Thus the small but non-zero eccentricity is likely a vestige of the initially high eccentricity that the planet obtained during some process that initiated high-eccentricity migration, a high eccentricity that was subsequently damped to the eccentricity we see today. Without a tighter constraint on the system's age, this is not conclusive. Also of interest is the high mass of TOI-628 b, which makes it one of only a few dozen known hot Jupiters with a mass $>$6 $M_{\rm J}$, and the most massive hot Jupiter found from \tess\ to date (see Figure \ref{fig:discussion}). 

\subsection{TOI-640 b}
The host star TOI-640 (TIC 147977348) is an F-star with a mass of $1.526^{+0.072}_{-0.079}$ $M_{\odot}$ and a radius of $2.082^{+0.064}_{-0.058}$ \rsun. The host star appears to be just transitioning off the main sequence into the subgiant branch as suggested by our measured $\log{g}$ = $3.987^{+0.030}_{-0.036}$ dex (cm s$^{-2}$) and corresponding tight age constraint within our global fit from the MIST evolutionary tracks of $1.99^{+0.55}_{-0.40}$ Gyr. It hosts a planetary companion, TOI-640 b, which is highly inflated ($R_{\rm P}$ = $1.771^{+0.060}_{-0.056}$ $R_{\rm J}$) Jupiter mass ($M_{\rm P}$ = $0.88\pm0.16$  $M_{\rm J}$) planet with a near integer orbital period of $5.0037775\pm0.0000048$ days. The orbit of the planet is consistent with circular, e = $0.050^{+0.054}_{-0.035}$. It is only the third hot Jupiter known with a highly inflated radius (R$_{\rm P}$>1.7) and on a period $>$5 days, joining KELT-12 b \citep{Stevens:2017} and Kepler-435 b \citep{Almenara:2015}. Interestingly, TOI-640 b is almost a twin of KELT-12 b, in that they are highly inflated Jupiter-mass planets on $\sim$5 day orbits around similar subgiant host stars. All three host stars in this regime are evolved, possibly suggesting that the inflation is a result of the host star's recent evolution \citep{Assef:2009, Spiegel:2012, Hartman:2016, Lopez:2016}. Similarly to KELT-12b and Kepler-435 b, we see no evidence of any significant eccentricity. 

\subsection{TOI-1333 b}
TOI-1333 (TIC 395171208) is a bright ($V$=9.49) evolved F-star with a mass of $M_{*}$ = $1.464^{+0.076}_{-0.079}$ $M_{\odot}$ and radius of $R_{*}$ = $1.925^{+0.064}_{-0.063}$ $R_{\odot}$. The star appears to be slightly evolved, as suggested by its $\log{g}$ of  $4.034^{+0.032}_{-0.033}$ dex (cm s$^{-2}$). As a result of its evolutionary stage, we estimate a relatively tight age constraint from the MIST evolutionary tracks of $2.33^{+0.71}_{-0.56}$ Gyr. Orbiting on a $4.720219\pm0.000011$ day period, the planetary companion TOI-1333 b has a radius of $1.396^{+0.056}_{-0.054}$ $R_{\rm J}$, a mass of $M_{\rm P}$ = $2.37\pm0.24$ $M_{\rm J}$, and an eccentricity that is consistent with circular (e = $0.073^{+0.092}_{-0.052}$). This is not surprising given that our derived circularization timescale, $1.29^{+0.37}_{-0.33}$ Gyr, is similar to the age of the system.

In the case of TOI-1333, we have measured a periodicity of $\sim$5.3 days from the ground-based and \tess\ photometry (we note that WASP identified a period 3$\times$ this). We have also measured a \vsini\ of 16.5$\pm$0.5 \kms. If the periodicity identified in the photometry indeed is the average rotation period of the host star, then we can estimate the inclination of the host star's rotation axis and compare it to the derived inclination of TOI-1333 b's orbit following the methodology presented in \citet{Masuda:2020}. Using the \texttt{EXOFASTv2} implementation of a Markov Chain Monte Carlo, we run a simple fit of the host star's rotational velocity and its projection onto our line of sight (\vsini) using the values from our global fit for Rstar ($1.925^{+0.064}_{-0.063}$ \rsun), the derived rotational period of TOI-1333 from \tess\ and KELT (5.3 days), and the \vsini\ from the TRES spectroscopy (14.2$\pm$0.5\kms) to calculate the inclination of TOI-1333's rotation axis (relative to our line of sight). The latitudes on the Sun that show star spots have a differential rotation on the surface of a few percent. Therefore, we place a 3\% error on the rotational velocity for this analysis. We require the same Gelman-Rubin statistic ($<$1.01) and independent draw ($>$1000) for convergence as the default for \texttt{EXOFASTv2}. We derive the inclination of the rotation axis to be 51.3$^{\circ{+3.5^\circ}}_{-3.3^\circ}$. From our global fit, TOI-1333 b has an inclination of $85.70^{\circ{+1.3^\circ}}_{-0.65^\circ}$, suggesting that the rotation axis of the star and the orbital plane are misaligned. TOI-1333 b is an excellent candidate to confirm this result through spin-orbit alignment ($\lambda$) measurements using the R-M or Doppler tomography techniques. The planet's orbit is also misaligned with the orbit of the wide binary companion TOI-1333 B, for which we measured an inclination of $125^{+18}_{-10}$ degrees from our LOFTI analysis. Interestingly, we do not detect a significant orbital eccentricity from our global fit for TOI-1333 b (though a small eccentricity is still possible) but this suggested misalignment might be a remnant left over from high eccentricity migration.  The likely bound companion at 470 au (see \S\ref{sec:AO}) could be responsible for Kozai-Lidov migration of the planet.

\subsection{TOI-1478 b}
TOI-1478 (TIC 409794137, $V$ = 10.81) is a Sun-like G-dwarf with radius of $R_{*}$ = $1.048^{+0.030}_{-0.029}$ $R_{\odot}$, mass of $M_{*}$ = $0.946^{+0.059}_{-0.041}$ $M_{\odot}$, and an age of $9.2^{+3.1}_{-3.9}$ Gyr. Orbiting TOI-1478 is a warm-Jupiter with a period of $10.180249\pm0.000015$ days, a radius of $R_{\rm P}$ = $1.060^{+0.040}_{-0.039}$ $R_{\rm J}$, a mass of $M_{\rm P}$ = $0.851^{+0.052}_{-0.047}$ $M_{\rm J}$, and resides in a circular orbit (e = $0.024^{+0.032}_{-0.017}$). TOI-1478 b is the longest period planet in our sample, and the planet and its host star (other than their orbital distances) resemble the Sun and Jupiter in mass and radius, possibly an example of an alternate outcome of our own Solar System. As a result of the long orbital period, the tidal forces on TOI-1478 b are too weak to have circularized the orbit. Therefore, the lack of a significant eccentricity could suggest a more dynamically quiescent migration history. 

\subsection{TOI-1601 b}
In the case of TOI-1601 b, our global model showed a clear bimodality in the posterior distribution of the host star's mass and age (See \S\ref{sec:TOI1601_bimodal}). This is likely due to the host star's evolutionary status, because the star sits on the HR diagram where isochrones cross, so the evolutionary state is ambiguous given the precision of our observations. and to account for this we extract two separate solutions, one for each peak in our posteriors. The higher host star mass solution, \mstar\ = $1.517^{+0.053}_{-0.049}$ \msun, has a higher probability of being correct at 66.7\%, so we adopt this solution for the discussion but both results are available in Table \ref{tab:exofast_stellar}. TOI-1601 (TIC 139375960, $V$ = 10.71) is an evolved subgiant ($\log{g}$ of $3.940^{+0.022}_{-0.025}$ dex (cm s$^{-2}$)) with a radius of $R_{*}$ = $2.186^{+0.074}_{-0.063}$ $R_{\odot}$. We estimate the age of the system within our fit to be $2.64^{+0.38}_{-0.39}$ Gyr. TOI-1601 b is a Jupiter mass planet ($0.99\pm0.11$ $M_{\rm J}$) that shows some inflation ($R_{\rm P}$ = $1.159^{+0.062}_{-0.059}$ $R_{\rm J}$) and a circular orbit (e = $0.037^{+0.045}_{-0.026}$), $5.331752\pm0.000011$ day orbit. The spectroscopic analysis of the TRES spectra of TOI-1601 shows some metal enhancement ([{\rm Fe/H}] = $0.316^{+0.072}_{-0.074}$).

\subsection{TESS's impact on Giant Planets}
While the primary goal of NASA's \TESS\ mission is to discover and measure the masses of small planets \citep{Ricker:2015}, \tess\ has already provided some valuable discoveries in the field of giant planets \citep[see, e.g.,][]{Huang:2020, Armstrong:2020, Vanderburg:2020}. Given the minimum $\sim$27 day baseline for any target, and the completeness in the sensitivity of space-based photometry to detect a hot-Jupiter transit, \tess\ provides the opportunity to obtain a near-complete sample of hot Jupiters \citep{Zhou:2019}. To date, \TESS\ has discovered 26 giant planets ($M_{\rm P}$ $>$ 0.4\mj), 16 of which have an orbital period $>$5.0 days (these numbers include the five systems presented in this paper and 2 additional systems from \citep{IkwutUkwa:2021}. For comparison, 36 hot Jupiters have been discovered with orbital periods $>$5.0 days from ground-based transit surveys (NASA Exoplanet Archive, \citealp{Akeson:2013}). 

If giant planets predominantly migrate through dynamical interactions, we may find evidence of this evolutionary history in the eccentricity distribution of hot Jupiters, specifically those that are dynamically young (where the circularization timescale by tidal forces is longer than the age of the system). Figure \ref{fig:discussion} shows the current distribution of giant planet eccentricities as a function of orbital period out to 16 days. Those 26 systems discovered by \tess\ are shown colored by their host star's effective temperature. Although this is not a homogeneous sample, since a variety of different analysis methods and assumptions were made within this population, there is a wider distribution of eccentricities for those systems with an orbital period $>$ 5 days, where tidal circularization timescales are longer \citep{Adams:2006}. Interestingly, of the 5 systems presented here, only TOI-628 b has a statistically significantly measured eccentricity (e = $0.072^{+0.021}_{-0.023}$), and is consistent with dynamically driven migration since its estimated age is less than the circularization time-scale of the orbit. Although the other systems show some non-zero eccentricities from our global fits, they are not statistically significant ($>$3$\sigma$) and could be a result of the Lucy-Sweeney bias \citep{Lucy:1971}. We also note that there is one very massive hot Jupiter in our sample, TOI-628 b (M$_P$ $6.33^{+0.29}_{-0.31}$ \mj), and it is the most massive hot Jupiter discovered to date by \tess\ (we note that \tess\ has discovered a few transiting brown dwarfs, \citealp{Jackman:2019, Subjak:2020, Carmichael2020A,Carmichael:2020B} and WD 1856+534 which has a mass limit $<$13.8 \mj, \citealp{Vanderburg:2020}). These massive Jupiters provide a great laboratory to study the effect of high gravity on the atmosphere of a gas giant, while studying the transition point between giant planets and brown dwarfs.

\section{Conclusion}
\label{sec:conclusion}
We present the discovery and characterization of five new giant planets (TOI-628, TOI-640 b, TOI-1333 b, TOI-1478 b, and TOI-1601 b) from NASA's \tess\ mission. These planets were discovered in the primary mission using the 30-minute cadenced, full frame images. Of the systems TOI-640 b, TOI-1333 b, and TOI-1601 b all orbit stars that appear to have just evolved off the main sequence entering the subgiant phase, as suggested by their estimated $\log{g}$ being under 4.1 dex (cm s$^{-2}$). None of the planets orbiting these subgiants appear to reside in a significantly eccentric orbits. TOI-628 b is the most massive hot Jupiter discovered by \tess\ ($M_{\rm P}$ = $6.33^{+0.29}_{-0.31}$ $M_{\rm J}$), and resides in an eccentric orbit that is consistent with dynamically driven migration. Another planet from this work, TOI-640 b, is one of the only highly inflated ($>1.7$ $R_{\rm J}$) hot Jupiters with an orbital period greater than 5 days. TOI-1478 b is the only planet in this sample with an orbital period $>$10 days, and it and its star are similar in size and mass to Jupiter and the Sun. All five planets orbit bright ($V<$10.7) host stars and significantly increase the sample of well-characterized, long period ($>$5 day) hot Jupiters, an area where NASA's \TESS\ mission should continue to provide a wealth of discoveries.

\acknowledgements

CZ is supported by a Dunlap Fellowship at the Dunlap Institute for Astronomy \& Astrophysics, funded through an endowment established by the Dunlap family and the University of Toronto. T.H. acknowledges support from the European Research Council under the Horizon 2020 Framework Program via the ERC Advanced Grant Origins 83 24 28. J.V.S. acknowledges funding from the European Research Council (ERC) under the European Union’s Horizon 2020 research and innovation programme (project Four Aces; grant agreement No. 724427). P. R. acknowledges support from NSF grant No. 1952545. R.B.\ acknowledges support from FONDECYT Project 11200751 and from CORFO project N$^\circ$14ENI2-26865. A.J.\, R.B.\, and M.H.\ acknowledge support from project IC120009 ``Millennium Institute of Astrophysics (MAS)'' of the Millenium Science Initiative, Chilean Ministry of Economy. D.J.S. acknowledges funding support from the Eberly Research Fellowship from The Pennsylvania State University Eberly College of Science. The Center for Exoplanets and Habitable Worlds is supported by the Pennsylvania State University, the Eberly College of Science, and the Pennsylvania Space Grant Consortium. K.K.M. gratefully acknowledges support from the NewYork CommunityTrust's Fund for Astrophysical Research. L.G. and A.G. are supported by NASA Massachusetts Space Grant Fellowships. E.W.G., M.E., and P.C. acknowledge support by Deutsche Forschungsgemeinschaft (DFG) grant  HA 3279/12-1  within the DFG Schwerpunkt SPP1992, Exploring the Diversity of Extrasolar Planets. B.S.G. was partially supported by the Thomas Jefferson Chair for Space Exploration at the Ohio State University. C.D. acknowledges support from the Hellman Fellows Fund and NASA XRP via grant 80NSSC20K0250.

We thank the CHIRON team members, including Todd Henry, Leonardo Paredes, Hodari James, Azmain Nisak,  Rodrigo Hinojosa, Roberto Aviles, Wei-Chun Jao and CTIO staffs, for their work in acquiring RVs with CHIRON at CTIO. This research has made use of SAO/NASA's Astrophysics Data System Bibliographic Services. This research has made use of the SIMBAD database, operated at CDS, Strasbourg, France. This work has made use of data from the European Space Agency (ESA) mission {\it Gaia} (\url{https://www.cosmos.esa.int/gaia}), processed by the {\it Gaia} Data Processing and Analysis Consortium (DPAC, \url{https://www.cosmos.esa.int/web/gaia/dpac/consortium}). Funding for the DPAC has been provided by national institutions, in particular the institutions participating in the {\it Gaia} Multilateral Agreement. This work makes use of observations from the LCO network. Based in part on observations obtained at the Southern Astrophysical Research (SOAR) telescope, which is a joint project of the Minist\'{e}rio da Ci\^{e}ncia, Tecnologia e Inova\c{c}\~{o}es (MCTI/LNA) do Brasil, the US National Science Foundation’s NOIRLab, the University of North Carolina at Chapel Hill (UNC), and Michigan State University (MSU).

Funding for the {\it TESS} mission is provided by NASA's Science Mission directorate. We acknowledge the use of public {\it TESS} Alert data from pipelines at the {\it TESS} Science Office and at the {\it TESS} Science Processing Operations Center. This research has made use of the NASA Exoplanet Archive and the Exoplanet Follow-up Observation Program website, which are operated by the California Institute of Technology, under contract with the National Aeronautics and Space Administration under the Exoplanet Exploration Program. This paper includes data collected by the {\it TESS} mission, which are publicly available from the Mikulski Archive for Space Telescopes (MAST). This paper includes observations obtained under Gemini program GN-2018B-LP-101. Resources supporting this work were provided by the NASA High-End Computing (HEC) Program through the NASA Advanced Supercomputing (NAS) Division at Ames Research Center for the production of the SPOC data products. This publication makes use of The Data \& Analysis Center for Exoplanets (DACE), which is a facility based at the University of Geneva (CH) dedicated to extrasolar planets data visualisation, exchange and analysis. DACE is a platform of the Swiss National Centre of Competence in Research (NCCR) PlanetS, federating the Swiss expertise in Exoplanet research. The DACE platform is available at \url{https://dace.unige.ch}.

Some of the data presented herein were obtained at the W. M. Keck Observatory, which is operated as a scientific partnership among the California Institute of Technology, the University of California and the National Aeronautics and Space Administration. The Observatory was made possible by the generous financial support of the W. M. Keck Foundation. The authors wish to recognize and acknowledge the very significant cultural role and reverence that the summit of Mauna Kea has always had within the indigenous Hawaiian community.  We are most fortunate to have the opportunity to conduct observations from this mountain.

\textsc{Minerva}-Australis is supported by Australian Research Council LIEF Grant LE160100001, Discovery Grant DP180100972, Mount Cuba Astronomical Foundation, and institutional partners University of Southern Queensland, UNSW Sydney, MIT, Nanjing University, George Mason University, University of Louisville, University of California Riverside, University of Florida, and The University of Texas at Austin. We respectfully acknowledge the traditional custodians of all lands throughout Australia, and recognise their continued cultural and spiritual connection to the land, waterways, cosmos, and community. We pay our deepest respects to all Elders, ancestors and descendants of the Giabal, Jarowair, and Kambuwal nations, upon whose lands the \textsc{Minerva}-Australis facility at Mt Kent is situated.

\software{EXOFASTv2 \citep{Eastman:2013, Eastman:2019}, AstroImageJ \citep{Collins:2017}, TAPIR \citep{Jensen:2013}, PEST Pipeline (\url{http://pestobservatory.com/the-pest-pipeline/}),LOFTI \citep{Pearce2020}, Isochrones package \citep{morton:2015}, QLP Pipeline \citep{Huang:QLP}, CERES \citep{Brahm:2017}}

\facilities{TESS, FLWO 1.5m (Tillinghast Reflector Echelle Spectrograph), 4.1-m Southern Astrophysical Research (SOAR), LCO 0.4m, LCO 1.0m, 2.2m telescope La Silla (Fiber-fed Extended Range Optical Spectrograph),  KECK (NIRC2), PALOMAR (PHARO), TESS, KELT, WASP, CTIO 1.5m (CHIRON), \textsc{Minerva}-Australis, GEMINI (NIRI)}

\bibliographystyle{apj}

\bibliography{refs}



\end{document}